\documentclass[superscriptaddress,aps,pra,twocolumn,showpacs]{revtex4}
\usepackage{bm}
\usepackage{ulem}
\usepackage{epsfig}
\usepackage{graphicx}
\usepackage{amssymb,amsmath,amsbsy,amsgen,amsfonts}    
\usepackage{dcolumn}
\usepackage{amsthm}
\usepackage{mathrsfs}
\usepackage{latexsym}
\usepackage{array}
\usepackage{color}
\usepackage{amstext}
\allowdisplaybreaks[1]
\usepackage{txfonts}
\usepackage{pifont}

\usepackage{epstopdf} %for texshop on mac

\newcommand{\bra}[1]{\langle#1\vert}
\newcommand{\ket}[1]{\vert#1\rangle}
\newcommand{\inner}[2]{\langle #1 |#2 \rangle}
\newcommand{\abs}[1]{\vert#1\vert}
\newcommand{\opd}[2]{{\hat{#1}_{#2}}^{\dagger}}
\newcommand{\op}[2]{\hat{#1}_{#2}}

\DeclareSymbolFont{symbols}{OMS}{cmsy}{m}{n}

\begin{document}
\title{Few Photon Transport in Many-Body Photonic Systems: A Scattering Approach}
\author{Changhyoup Lee}
\email{changdolli@gmail.com}
\affiliation{Centre for Quantum Technologies, National University of Singapore, 3 Science Drive 2, Singapore 117543}
\author{Changsuk Noh} 
\email{undefying@gmail.com}
\affiliation{Centre for Quantum Technologies, National University of Singapore, 3 Science Drive 2, Singapore 117543}
\author{Nikolaos Schetakis}
\affiliation{School of Electronic and Computer Engineering, Technical University of Crete, Chania, Greece 73100}
\author{Dimitris G. Angelakis}
\email{dimitris.angelakis@gmail.com}
\affiliation{Centre for Quantum Technologies, National University of Singapore, 3 Science Drive 2, Singapore 117543}
\affiliation{School of Electronic and Computer Engineering, Technical University of Crete, Chania, Greece 73100}
\date{\today}

\begin{abstract}
We study the quantum transport of multi-photon Fock states in one-dimensional  Bose-Hubbard lattices implemented in QED cavity arrays (QCAs). We propose an optical scheme to probe the underlying many-body states of the system by analyzing the properties of the transmitted light using scattering theory. To this end, we employ the Lippmann-Schwinger formalism within which an analytical form of the scattering matrix can be found. The latter is evaluated explicitly for the two particle/photon-two site case using which we study the resonance properties of two-photon scattering, as well as the scattering probabilities and the second-order intensity correlations of the transmitted light.  The results indicate that the underlying structure of the many-body states of the model in question can be directly inferred from the physical properties of the transported photons in its QCA realization. We find that a fully-resonant two-photon scattering scenario allows a faithful characterization of the underlying many-body states, unlike in the coherent driving scenario usually employed in quantum Master equation treatments. The effects of losses in the cavities, as well as the incoming photons' pulse shapes and initial correlations are studied and analyzed. Our method is general and can be applied to probe the structure of any many-body bosonic models amenable to a QCA implementation including the Jaynes-Cummings-Hubbard, the extended Bose-Hubbard as well as a whole range of spin models.
\end{abstract}


\maketitle
%%%%%%%%%%%%%%%%%%%%%%%%%%%%%%%%%%%%%%%%%%%%%%%%%%%%%%%%%%%%%%%%%%%%%
%%%%%%%%%%%%%%%%%%%%%%%%%%%%%%%%%%%%%%%%%%%%%%%%%%%%%%%%%%%%%%%%%%%%%
\section{Introduction} 
Recent advances in quantum nonlinear optics and circuit QED systems \cite{Houck12,Chang14} have allowed the engineering of photon-photon interaction to the extent that strongly interacting photons have started to be  considered as a potential platform to simulate many-body phenomena \cite{Hartmann08,Tomadin10a,Schmidt13,Carusotto13}. Early proposals discussed the possibility to realise strongly correlated states of photons and polaritons in coupled QED cavity arrays (QCAs) \cite{Hartmann06,Greentree06,Angelakis07}. Their natural advantage in local control and design, and possibility to probe out-of-equilibrium phenomena in driven dissipative regimes, allowed QCA-based approaches to  complement the efforts towards viable quantum simulators \cite{Carusotto09, Tomadin10, Hartmann10, Nunnenkamp11,Nissen12,Grujic12,Grujic13,Jin13,Biella15}. Experimentally, in spite of various challenges,  progress has been recently made with small scale QCAs  successfully fabricated in semiconductor and superconductor  based  set-ups \cite{ Abbarchi13, Raftery14, Eichler14}. Strongly interacting photons have also been created in Rydberg media \cite{Peyronel12}.
\begin{figure}
\centering
\includegraphics[width=0.5\textwidth]{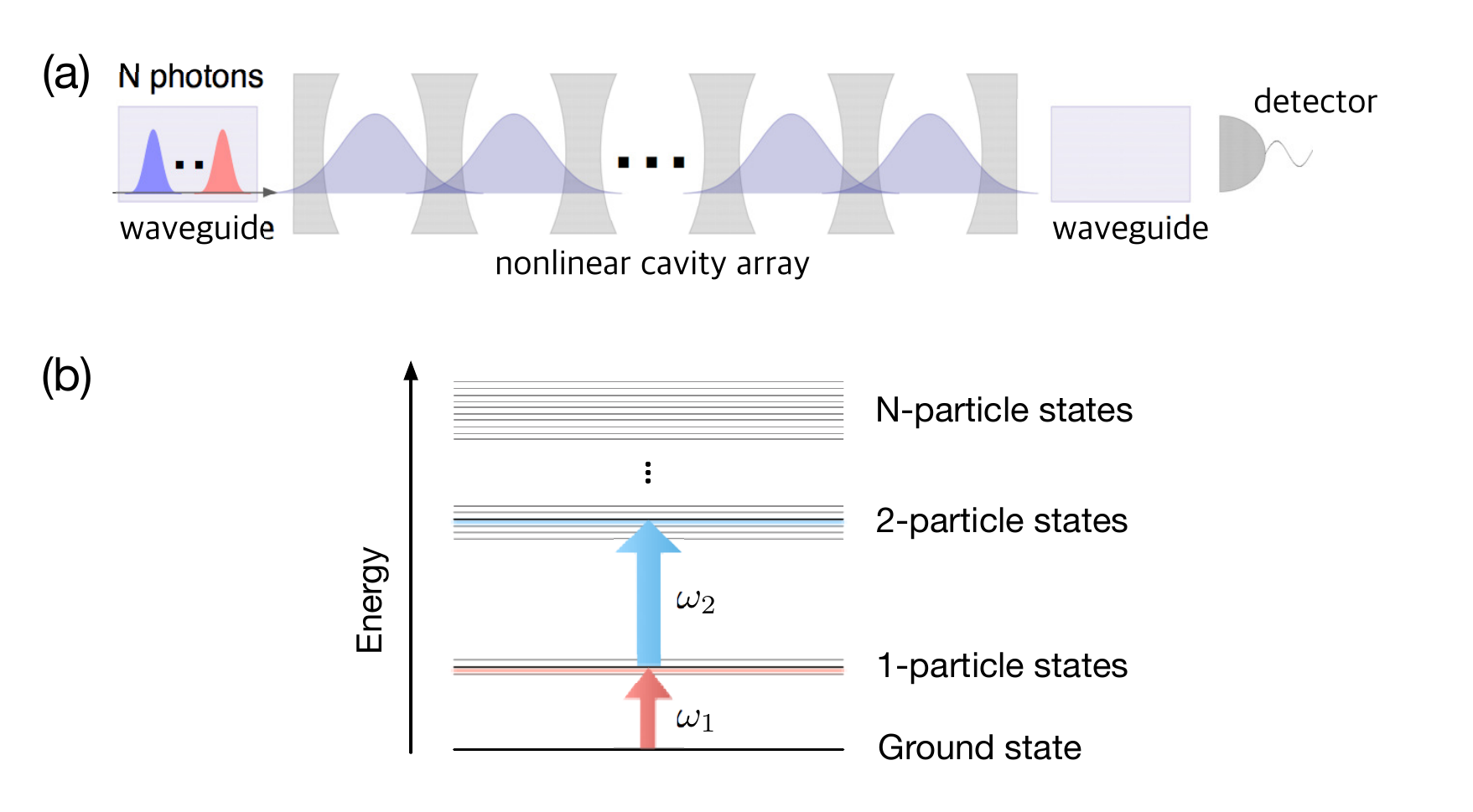}
\caption{
(a) Proposed method to probe the structure of bosonic many-body models as implemented QCA simulators.
Photons traveling in the left waveguide are injected into the array and are transported through the device to the right waveguide. In this work, the QCA  is assumed to realize the Bose-Hubbard model but other models such as the Jaynes-Cummings-Hubbard, spin models, or the extended Bose-Hubbard can also be realized \cite{Angelakis07,Jin13}.  The injected photons scan through the many-body eigenstates of the simulated model and if they are fully resonant to the many-body states as illustrated in (b), the full information of the relevant states is mapped out faithfully in the output spectra and correlation functions.
}
\label{system} 
\end{figure}

A QCA, beyond its many-body character, is inherently a (quantum) optical system, thus is naturally probed by light scattering \cite{Carmichael}. %
Performing quantum measurements on the output (transported/scattered) light, one obtains information about the underlying properties of the system \cite{footnote}. 
In the study of QCA simulators, the driving source has so far mostly been taken to be a coherent field of light described within a quantum Master equation formalism. The latter approach, although successfully captures the open nature of the system, is often limited to coherent-light drives (recently a method to derive a master equation for the Fock-state input has been found in \cite{Baragiola12}). This semiclassical treatment misses in our opinion an important regime of input quantum particles being transported in the system. How does a QCA many body simulator react to general quantum input fields? Can we collect information on the states of the many-body models simulated by studying the transported/scattered quantum particles (photons) from a QCA?

To answer this question, we employ the Lippmann-Schwinger formalism, whose use in quantum optical systems was pioneered by Shen and Fan \cite{Shen07} and 
led to numerous further developments  \cite{Shen07a,Shen09,Nishino09,Fan10,Liao10,Zheng10,Zheng11,Shi11, Shi13,Rephaeli12,Xu13,Laakso14,Zheng13}.
In the context of quantum simulations of many-body phenomena, using an $N$-photon Fock state as the input field for an $N$-site cavity array seems promising. As a first step towards this goal, we examine the process of scattering two photons on an array of coupled Kerr nonlinear resonators whose dynamics are described by the Bose-Hubbard model. We first evaluate the scattering matrix analytically for the case of two resonators coupled to input and output waveguides, and then use it to calculate the scattering probabilities and the second-order correlations between the scattered photons. 
The results indicate that the structure of the correlated many-body states is more clearly reflected in the scattered light fields when the individual input photon/particle energies are fully resonant with the corresponding eigenstates (see Fig.~\ref{system}).

%%%%%%%%%%%%%%%%%%%%%%%%%%%%%%%%%%%%%%%%%%%%%%%%%%%%%%%%%%%%%%%%%%%%%
%%%%%%%%%%%%%%%%%%%%%%%%%%%%%%%%%%%%%%%%%%%%%%%%%%%%%%%%%%%%%%%%%%%%%
\section{Few-Photon Transport}
%\section{System Hamiltonian}
Consider a one-dimensional array of $N$ coupled nonlinear cavities, where the cavities at both ends are coupled to waveguides supporting propagating photons as shown in Fig.~\ref{system}(a). The system is described by the Hamiltonian, 
\begin{eqnarray*}
\hat{H}_{\rm tot}=\hat{H}_{\rm wg}+\hat{H}_{\rm cc}+\hat{H}_{\rm wc},
\end{eqnarray*}
where
\begin{eqnarray}
&&\hat{H}_{\rm wg}=\hbar \int_{-\infty}^{\infty}dx \Big(-i v_{g}\opd{c}{L}(x) \frac{\partial}{\partial x}\op{c}{L}(x) \Big)\nonumber\\
&&~~~~~~~~~~~~~~~~~~~~~~~~~~+\hbar \int_{-\infty}^{\infty}dy \Big(-i v_{g} \opd{c}{R}(y) \frac{\partial}{\partial y}\op{c}{R}(y) \Big), \nonumber\\
&&\hat{H}_{\rm cc}=\hbar \sum_{j=1}^{N}\Big( \omega_{j}\opd{a}{j}\op{a}{j} + U_{j} \opd{a}{j} \opd{a}{j} \op{a}{j}\op{a}{j} \Big) \nonumber \\
&&~~~~~~~~~~~~~~~~~~~~~~~~~~+\hbar \sum_{j=1}^{N-1}J(\opd{a}{j}\op{a}{j+1}+\op{a}{j}\opd{a}{j+1}), \nonumber\\
&&\hat{H}_{\rm wc}=\hbar \int_{-\infty}^{\infty}dx V_{1}\delta(x)\Big(\opd{c}{L}(x)\op{a}{1}+\op{c}{L}(x)\opd{a}{1}\Big)  \nonumber\\
&&~~~~~~~~~~~~~~~~~~~~~~~~~+\hbar \int_{-\infty}^{\infty}dy V_{2}\delta(y)\Big(\opd{c}{R}(y)\op{a}{N}+\op{c}{R}(y)\opd{a}{N} \Big). \nonumber
\end{eqnarray}
$\hat{H}_{\rm wg}$ describes the propagation of photons in the waveguides with group velocity $v_g$, where $\opd{c}{L}(x<0)$ $\left ( \opd{c}{L}(x>0) \right)$ and $\opd{c}{R}(y<0)$ $\left (\opd{c}{R}(y>0) \right )$ are the creation operators for an incoming (outgoing) photon in the left and right waveguides, respectively. 
$\hat{H}_{\rm cc}$ describes the coupled cavity system, where the bosonic operator $\opd{a}{j}$ annihilates a photon in the $j$th cavity which has the resonant frequency $\omega_{j}$ and nonlinearity $U_{j}$. The photon hopping rate between the cavities is given by $J$. $\hat{H}_{\rm wc}$ describes the coupling between the waveguides to the adjacent cavities, with coupling strengths $V_{1}$ and $V_{2}$. From here on, we set $v_{g} = \hbar = 1$.
\begin{figure}
\centering
\includegraphics[width=8.5cm]{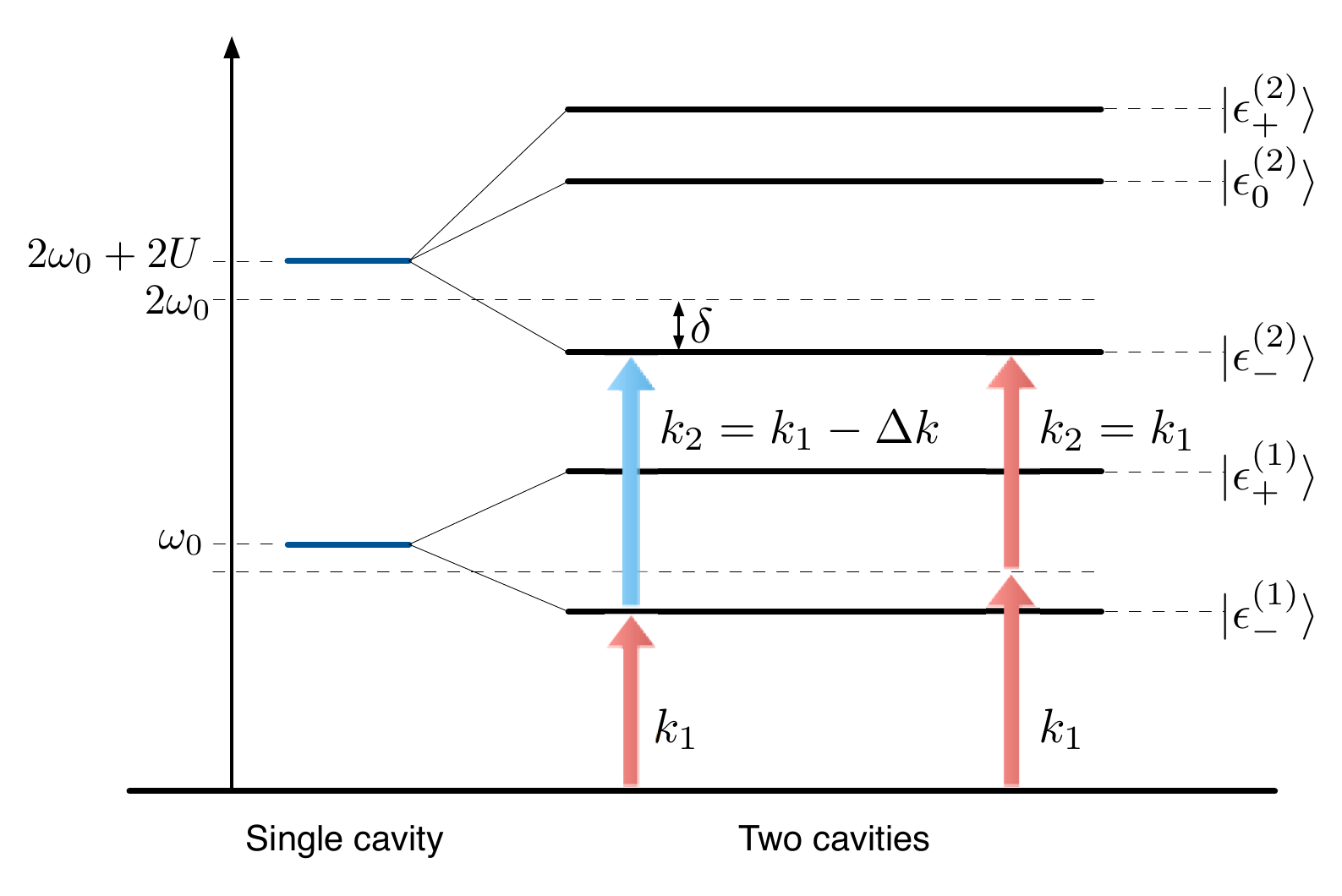}
\caption{
Energy level diagram of the two-site Bose-Hubbard QCA, where the bare-cavity energies and the coupled mode energies of the cavities are shown.One of the resonant two-photon excitation paths satisfying Eq.~(\ref{res.cond}) is illustrated by the arrows on the left, while the off-resonant path for identical input photons is shown on the right.
%\textcolor{red}{All the resonant transitions are encoded in the scattering matrix. One of the resonant two-photon excitation paths, satisfying Eq.~(\ref{res.cond}), is illustrated by arrows on the left side, whereas the off-resonant excitation path is depicted on the right side, where the virtual single-particle excitation is involved.}
}
\label{transition} 
\end{figure}
%%%%%%%%%%%%%%%%%%%%%%%%%%%%%%%%%%%%%%%%%%%%%%%%%%%%%%%%%%%%%%%%%%%%%
%%%%%%%%%%%%%%%%%%%%%%%%%%%%%%%%%%%%%%%%%%%%%%%%%%%%%%%%%%%%%%%%%%%%%
%\section{Scattering matrix of a QCA}
To analyse the properties of the scattered photons, we analytically find the two-photon scattering matrix ${\bf S}^{(2)}$ within the Lippmann-Schwinger formalism (for a formal definition of the scattering matrix and a detailed derivation, see Appendix~\ref{App_scattering_eigenstate}):
\begin{align}
_{LL}\bra{p_{1},p_{2}} &{\bf S}^{(2)}  \ket{k_{1},k_{2}} = S_{LL}  \delta(k_{1}+k_{2}-p_{1}-p_{2}) \nonumber \\ &+ \big( r_{k_{1}} r_{k_{2}} \delta(k_{1}-p_{1})\delta(k_{2}-p_{2}) +(k_1 \leftrightarrow k_2)\big), \label{scatering_matrix_element_LL}   \\
 _{LR}\bra{p_{1},p_{2}} &{\bf S}^{(2)}  \ket{k_{1},k_{2}} =S_{LR}  \delta(k_{1}+k_{2}-p_{1}-p_{2}) \nonumber \\ &+ \big( r_{k_{1}} t_{k_{2}} \delta(k_{1}-p_{1})\delta(k_{2}-p_{2})+ (k_1 \leftrightarrow k_2) \big), \label{scatering_matrix_element_LR} \\
 _{RR}\bra{p_{1},p_{2}} &{\bf S}^{(2)}  \ket{k_{1},k_{2}} = S_{RR}  \delta(k_{1}+k_{2}-p_{1}-p_{2}) \nonumber \\ &+ \big ( t_{k_{1}} t_{k_{2}}\delta(k_{1}-p_{1})\delta(k_{2}-p_{2}) + (k_1 \leftrightarrow k_2) \big), \label{scatering_matrix_element_RR}
\end{align}
where we have used $k_i$ ($p_i$) to denote the input (output) momenta. The subscripts $LL$, $LR$, and $RR$ refer to which waveguide the two output photons have scattered to, e.g., $RR$ means that two photons are in the right waveguide. $r_k$ and $t_k$ are the single-photon reflection and transmission coefficients, respectively. The second lines on the right-hand side of the equations describe independent single-photon scattering events, whereas the first lines describe the contributions due to the nonlinearity present in the cavity array,
i.e., $S_{LL}$, $S_{LR}$, and $S_{RR}$ vanish when $U_{1}=U_{2}=0$.

\begin{figure*}
\centering
\includegraphics[width=13cm]{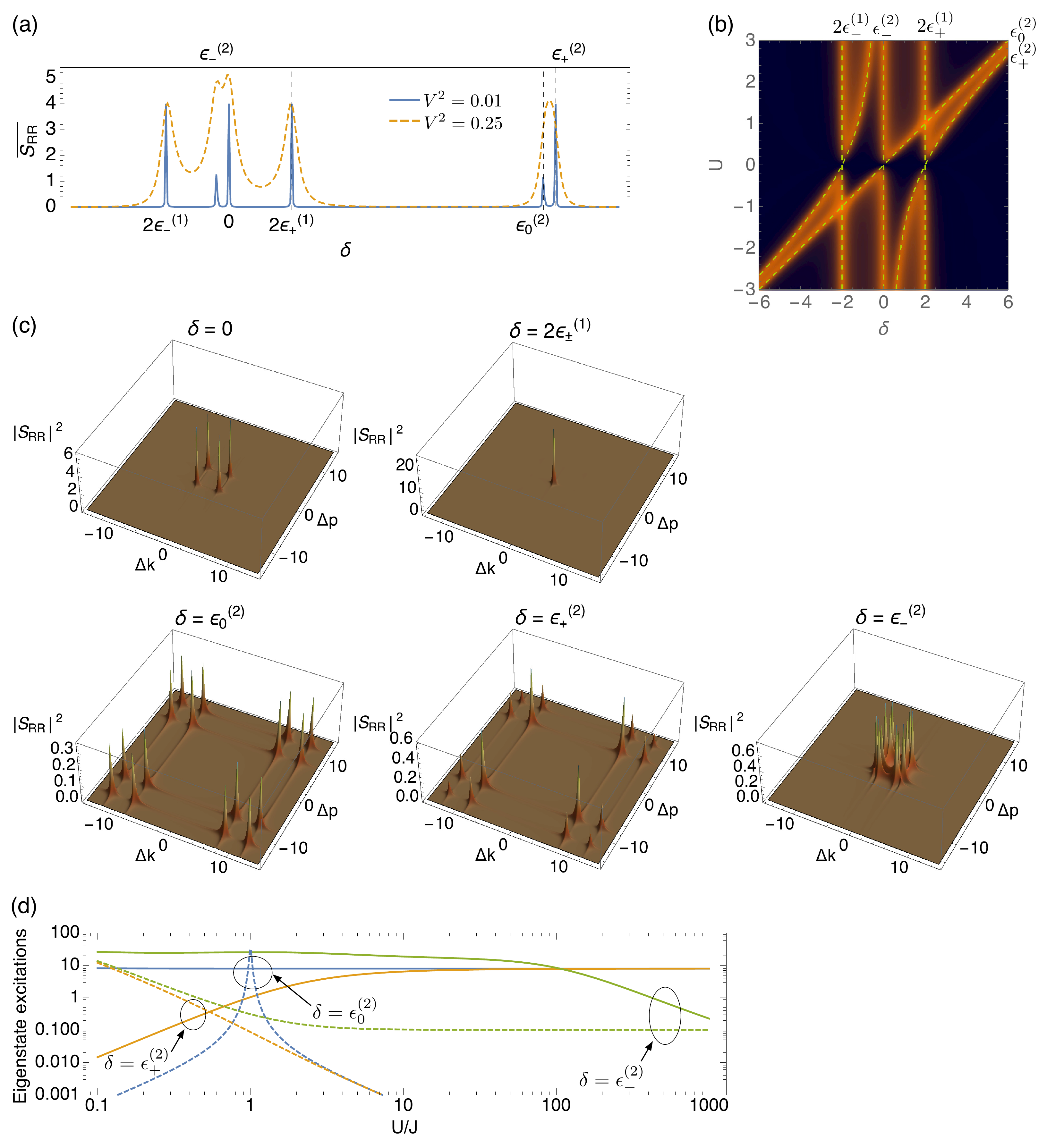}
\caption{
Bound state contribution in the scattering matrix element, $\overline{S_{RR}}$, is shown as a function of $\delta$, for $V^{2}=0.01$ (blue solid) and $V^{2}=0.25$ (orange dashed) in (a), and with detuning of total incident energy and nonlinearity for $V^{2}=0.25$ in (b).
Panel (c) shows the resonant conditions of $\Delta k(\Delta p)$ in $\abs{S_{RR}}^{2}$ for $\delta=0, 2\epsilon^{(1)}_{\pm}$, and $\epsilon^{(2)}_{0,\pm}$ for $U=5$ and $V^{2}=0.25$. 
Panel (d) presents the respective eigenstate excitation amplitudes at the corresponding two-photon energy resonances with increasing $U/J$ when $\Delta k=\delta-2\epsilon_{-}^{(1)}$ (solid) and $\Delta k=0$ (dashed) for $V^{2}=0.04$.
All units are defined with respect to $J$. 
%Note the scattering matrix $\abs{S_{RR}}^{2}$ in terms of $\delta$ and $\Delta k (\Delta p)$ clearly reads the resonant condition of Eq.~(\ref{res.cond}), and the desired eigenstates are more efficiently excited when Eq.~(\ref{res.cond}) is satisfied, implying the fully-resonant case more strongly connects the many-body correlated states of QCA with the output photons. 
Panel (c) clearly depicts the resonance condition written in Eq.~(\ref{res.cond}) while panel (d) shows how the desired eigenstates are more efficiently excited when this condition is met (solid curves) as opposed to the off-resonant case (dashed curves). 
}
\label{SMresonance} 
\end{figure*}

We now focus on the experimentally relevant case of two resonators \cite{Abbarchi13, Raftery14, Eichler14}, and assume for simplicity  $\omega_{1}$=$\omega_{2}$=$\omega_{0}$, $U_{1}$=$U_{2}$=$U$, and $V_{1}$=$V_{2}$=$V$. At this point it is useful to define the total energy $k_{1}+k_{2} = p_{1}+ p_{2}$ as $2\omega_{0}+\delta$ and the relative energy as $\Delta k =k_{1}-k_{2}$ and $\Delta p =p_{1}-p_{2}$. The eigenenergies of the system in the one-particle manifold are $\omega_{0} + \epsilon_\pm^{(1)}$ where $\epsilon_\pm^{(1)}\equiv \pm J$, and the two-particle excitation subspace is composed of $2\omega_{0}+ \epsilon_{0,\pm}^{(2)}$ with $\epsilon_{0}^{(2)}\equiv 2U$ and $\epsilon_{\pm}^{(2)}\equiv U\pm \sqrt{4J^{2}+U^{2}}$ corresponding, respectively, to the eigenstates
\begin{eqnarray}
\ket{2_{0}} &\sim& \ket{20} - \ket{02}, \nonumber\\
\ket{2_{\pm}} &\sim& \ket{20}+\ket{02} - \frac{U\mp \sqrt{4J^{2}+U^{2}}}{\sqrt{2}J}\ket{11},\nonumber
\end{eqnarray}
where $\ket{jk}=\frac{1}{\sqrt{j! k!}}(\opd{a}{1})^{j}(\opd{a}{2})^{k}\ket{0}$.
Here, $\ket{2_{-}}$ becomes the unit-filled ground state $\ket{11}$ in the limit of $U\rightarrow\infty$.
In Eqs.~(\ref{scatering_matrix_element_LL})-(\ref{scatering_matrix_element_RR}), the bound-terms $S_{LL}, S_{LR}$, and $S_{RR}$ have resonances at 
\begin{equation}
\vert \Delta k\vert (\vert\Delta p\vert) =\vert2\epsilon_{\pm}^{(1)} - \delta \vert 
\label{res.cond}
\end{equation} 
for $\delta = 0, 2\epsilon_{\pm}^{(1)}$, and $\epsilon^{(2)}_{0,\pm}$, implying that the bound-term contributions are significant only if one of the input or output photons is resonant with one of the single-photon eigenstates as illustrated in Fig.~\ref{transition}.

First we discuss the resonance structure of the scattering matrix. To show an example of how the bound terms behave, we depict $\overline{S_{RR}} = \int d\Delta k d\Delta p \vert S_{RR} \vert^{2}$ as a function of $\delta$ in Fig.~\ref{SMresonance}(a). When the waveguide-cavity coupling strength is weak (blue solid curve, $V^{2}=0.01$) we find that the resonant peaks at $\delta = 0, 2\epsilon^{(1)}_{\pm}$, and $\epsilon^{(2)}_{0,\pm}$ are clearly distinguished, whereas for a higher coupling strength (orange dashed curve, $V^{2}=0.25$) resonances get broaden such that finer details are washed out. General resonant behaviour of $\overline{S_{RR}}$ over $\delta$ and $U$ is also depicted in Fig.~\ref{SMresonance}(b), which shows that the bound-terms have the resonances at $\delta = 0, 2\epsilon^{(1)}_{\pm}$, and $\epsilon^{(2)}_{0,\pm}$ for any value of $U$.
Furthermore, $\abs{S_{RR}}^{2}$ is analyzed as a function of $\Delta k$ and $\Delta p$ for each resonant $\delta$ in Fig.~\ref{SMresonance}(c), where the resonant condition of Eq.~(\ref{res.cond}) for $\Delta k$ $(\Delta p)$ is clearly seen. 
The first two cases ($\delta=0$ and $2\epsilon^{(1)}_{\pm}$) correspond to when each photon is resonant to a state belonging to the single excitation manifold, while the rest ($\delta=\epsilon_{0,\pm}^{(2)}$) correspond to when one photon has either $\epsilon^{(1)}_{\pm}$, and the other has $\epsilon^{(2)}_{0,\pm}-\epsilon^{(1)}_{\pm}$.
Similar resonant mechanisms have been observed in other systems such as a waveguide coupled to a cavity embedded in a two-level system \cite{Shi11} or a waveguide coupled to a whispering-galley resonator containing an atom \cite{Shi13}.

Throughout this work, we will consider two types of input states: 1) two photons satisfying the resonance condition (\ref{res.cond}), where for simplicity one of the input photons is assumed to have the energy $\epsilon^{(1)}_-$, i.e., $\Delta k =\delta-2\epsilon_{-}^{(1)}$ with $\delta = \epsilon_{0,\pm}^{(2)}$ (see arrows on the left side of Fig.~\ref{transition}); 
2) two photons satisfying the two-photon resonance condition while having the same energy, i.e., $\Delta k =0$ with $\delta = \epsilon_{0,\pm}^{(2)}$ (see arrows on the right side of Fig.~\ref{transition}). 
Later, we will show that, within the long input pulse regime, the second-order intensity correlations in the latter case is directly proportional to that in the coherent driving scenario.
Figure \ref{SMresonance}(d) shows the (unnormalised) two-photon eigenstate excitation amplitudes directly involved in two-photon scattering constructed from the coefficients ($e_{11}$, $e_{12}$ and $e_{22}$) of the two-photon scattering eigenstate given in Appendix~\ref{App_two_photon_scattering}. 
We see that when driven by the respective two-photon eigenenergies (three circles), the fully-resonant case (solid curves) generally excites the desired eigenstates more efficiently than the identical-photon input case (dashed curves). Exceptions only occur in two regimes: 1) near the linear regime for $\delta = \epsilon_+^{(2)}$, where the $\Delta k = 0$ hits the higher harmonic ladder; 2) near $U/J = 1$ for $\delta = \epsilon_0^{(2)}$, where the two-photon energy becomes twice the single photon eigenenergy $\epsilon_{+}^{(1)}$.
The fully-resonant photon scattering scenario therefore promises more efficient probe transmission spectroscopy of the multi-photon eigenstates. We will show this by explicitly calculating the scattering probabilities. We also calculate the second-order intensity correlations to further characterise the scattered light and connect the observed behaviour with the underlying states of the QCA.

\begin{figure*}[t]
\centering
\includegraphics[width=17cm]{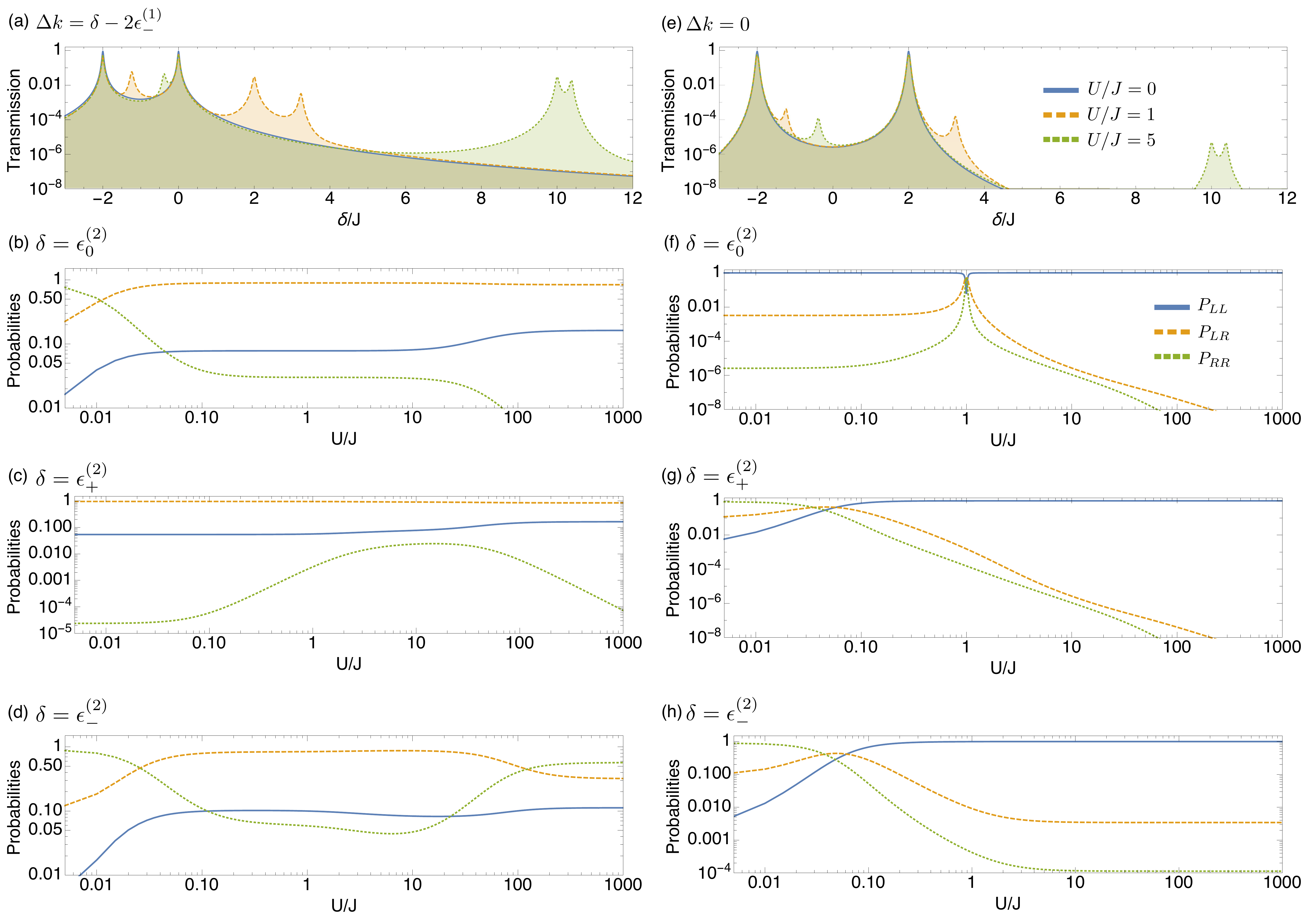}
\caption{
Left- and right-hand columns : $\Delta k =\delta-2\epsilon_{-}^{(1)}$ and $\Delta k =0$. 
Two-photon transmission ($P_{RR}$) is shown as a function of the two-photon detuning $\delta/J$ for different photon-photon interaction strengths $U/J=0,1,5$ in (a) and (e). The probabilities, $P_{LL}, P_{LR}$, and $P_{RR}$, are also shown as a function of $U/J$ for two-photon eigenenergies $\delta=\epsilon^{(2)}_{0,\pm}$ in (b)-(d) and (f)-(h).
Weak waveguide-cavity coupling and the narrow bandwidth of initial photons are assumed: $V^{2}/J=0.04$ and $\sigma/J=0.005$.
Note the difference in the behaviour of scattering probabilities as a function of $U/J$  when two-particle states are probed fully resonantly with different energy photons via the one-particle manifold (left column) compared to the case where a virtual (off-resonant) one-photon absorption is required (right column). In addition, in the former case, transmission is generally significantly larger which makes this approach experimentally more efficient (see text for more details).}
\label{probabilities} 
\end{figure*}

\section{Signatures of many-body states in transmission spectra}
%%%%%%%%%%%%%%%%%%%%%%%%%%%%%%%%%%%%%%%%%%%%%%%%%%%%%%%%%%%%%%%%%%%%%
%%%%%%%%%%%%%%%%%%%%%%%%%%%%%%%%%%%%%%%%%%%%%%%%%%%%%%%%%%%%%%%%%%%%%
%\subsection{Two-photon input state}
In the momentum space, a general two-photon initial state is given by $\ket{2_{\{ \xi \}}} = \frac{1}{\sqrt{M_{2}}} \opd{c}{\xi_{1}}\opd{c}{\xi_{2}}\ket{0}$, where the normalisation factor $M_{2}=1+ \big\vert \int dk\xi_{1}(k)\xi_{2}(k) \big\vert^{2}$ is associated with the overlap of the momentum distributions $\xi_i(k)$ and the continuous-mode creation operator is given by $\hat{c}_{\xi}^{\dagger} = \int dk \xi(k) \hat{c}_{L}^{\dagger}(k) $ with $\int dk \abs{\xi(k)}^{2}=1$.
The output state is then calculated from the scattering matrix as follows: 
\begin{eqnarray}
\ket{{\rm out}_{\{ \xi \}}^{(2)}} = {\bf S}^{(2)} \ket{2_{\{ \xi \}}} = \ket{{\rm out}_{\{ \xi \}}^{(2)}}_{LL} + \ket{{\rm out}_{\{ \xi \}}^{(2)}}_{LR}+ \ket{{\rm out}_{\{ \xi \}}^{(2)}}_{RR},\nonumber
\end{eqnarray}
where $\ket{{\rm out}_{\{ \xi \}}^{(2)}}_{\rm s_{1}s_{2}} = \int dq_{1}dq_{2} \frac{1}{\sqrt{M_{2}}}  \xi_{1}(q_{1})\xi_{2}(q_{2}) \ket{\phi_{\rm out}^{(2)}}_{\rm s_{1}s_{2}}$, for $(s_{1},s_{2}) \in \{ L, R \}$, where $\ket{\phi_{\rm out}^{(2)}}_{LL}$, $\ket{\phi_{\rm out}^{(2)}}_{LR}$ and $\ket{\phi_{\rm out}^{(2)}}_{RR}$ represent the two-photon wave functions associated with Eqs.~(\ref{scatering_matrix_element_LL}), (\ref{scatering_matrix_element_LR}), and (\ref{scatering_matrix_element_RR}), respectively (see Appendix~\ref{App_two_photon_scattering}). 
We assume the momentum distribution to have a narrow Gaussian profile for simplicity, i.e., $\xi_{j}(q)=\frac{1}{(2\pi \sigma^{2})^{1/4}} {\rm exp}\Big( -\frac{(q-k_{j})^{2}}{4\sigma^{2}}\Big)$, where $\xi_{j}$ is narrowly peaked around $k_j$. 
Given a narrow enough bandwidth with respect to the effective cavity linewidth, $\propto V^{2}$, effects of the pulse shape are very small as presented in Appendix~\ref{App_pulse_shape}--quantitatively similar results are obtained for both the Lorentzian and `rising' pulse profiles.
We note that recent developments in the pulse-shaping techniques makes our photon scattering scenario experimentally feasible \cite{gaussian, rising}. 

%%%%%%%%%%%%%%%%%%%%%%%%%%%%%%%%%%%%%%%%%%%%%%%%%%%%%%%%%%%%%%%%%%%%%
%%%%%%%%%%%%%%%%%%%%%%%%%%%%%%%%%%%%%%%%%%%%%%%%%%%%%%%%%%%%%%%%%%%%%
\subsection{Scattering probabilities}
Using the above initial state, we first consider the scattering probabilities defined as,
\begin{eqnarray}
P_{LL} &=& \int dp_{1}dp_{2} \frac{1}{2} \abs{\inner{p_{1},p_{2}}{{\rm out}_{\{ \xi \}}^{(2)}}_{LL} }^{2},\nonumber\\
P_{LR} &=& \int dp_{1}dp_{2} \abs{\inner{p_{1},p_{2}}{{\rm out}_{\{ \xi \}}^{(2)}}_{LR} }^{2}, \nonumber \\
P_{RR} &=& \int dp_{1}dp_{2}  \frac{1}{2} \abs{\inner{p_{1},p_{2}}{{\rm out}_{\{ \xi \}}^{(2)}}_{RR} }^{2}, \nonumber
\end{eqnarray}
as in \cite{Zheng10}. Figure~\ref{probabilities} depicts them as functions of the total energy $\delta/J$ (top row), or of the photon-photon interaction strength $U/J$ (lower rows). 
Left-hand column displays the fully-resonant case (see Eq.~(\ref{res.cond})) where one photon has the energy $\epsilon_{-}^{(1)}$ and the other has the energy $\delta -\epsilon_{-}^{(1)}$, whereas the right-hand column displays the results when $\Delta k = 0$.
In Figs.~\ref{probabilities}(a) and (e), we plot the two-photon transmission probability ($=P_{RR}$) for different values of interaction strengths ($U/J=0,1,5$).
In the linear case, there are transmission peaks when each photon is resonant to the linear mode of the coupled cavities. As one increases the nonlinearity, peaks start to form at the correlated two-particle eigenstates of the coupled nonlinear cavities.

Note that the transmission probabilities are significantly larger in the fully-resonant cases compared to the $\Delta k=0$ cases, in which the two-photon transmission requires a virtual (off-resonant) one-photon absorption. 
This indicates that the fully-resonant Fock-state transport scheme has an advantage over the identical-photon transport case in detecting two-photon transmission through the multi-particle correlated states of the QCA. 
In turn, this means that the two-photon scattering scenario performs better than the coherent driving case because: 1) the two-photons necessarily have the same energy in the latter and 2) the probability of finding two photons in a coherent state $\ket{\alpha}$ goes as $\abs{\alpha}^4 \ll 1$ in the weak-field limit.

In Figs.~\ref{probabilities}(b)-(d) and (f)-(h), the scattering probabilities at the resonances $\delta=\epsilon_{0,\pm}^{(2)}$ are further investigated as functions of $U/J$. 
We first note that over a wide region of $U/J$, except for the cases that coincide with the single photon resonances, $P_{LR}\approx 1$ for $\Delta k=\delta-2\epsilon_{-}^{(1)}$ (left hand column), while $P_{LL}\approx 1$ for $\Delta k =0$. This is due to the fact that one of the two photons is always resonant to the (lower) single energy state in the fully-resonant case, while neither photon is resonant in the $\Delta k=0$ cases. 
The figure also hints that the probabilities at $\delta=\epsilon_{0}^{(2)}$ and $\delta=\epsilon_{+}^{(2)}$ approach the same value above $U/J\sim 20$. This is due to the fact that above this value of $U/J$, the two states are no longer distinguishable because of their energy broadening ($V^{2}/J=0.04$). The interference between the corresponding eigenstates, $\ket{2_{0}}$ and $\ket{2_{+}}$, induces the little shift observed in the scattering probabilities. Similarly, in $\delta=\epsilon_-^{(2)}$ case, the energy of one of the photons approach $\epsilon_+^{(1)}$ within the decay bandwidth, resulting in larger two-photon transmission probability with increasing $U/J$. Effects of this kind are absent when $\Delta k = 0$.  

%%%%%%%%%%%%%%%%%%%%%%%%%%%%%%%%%%%%%%%%%%%%%%%%%%%%%%%%%%%%%%%%%%%%%
%%%%%%%%%%%%%%%%%%%%%%%%%%%%%%%%%%%%%%%%%%%%%%%%%%%%%%%%%%%%%%%%%%%%%
\subsection{Intensity-intensity correlations}
The scattering probabilities reveal the presence of the multi-photon correlated states, but no information about the actual correlations is given. For the latter, one may employ the second-order correlation function between positions $z_{1}$ and $z_{2}$:
$g_{s_{1}s_{2}}^{(2)} (z_{1},z_{2})= \frac{\bra{{\rm out}^{(2)}_{\{\xi\}}} \opd{c}{s_{1}}(z_{1})\opd{c}{s_{2}}(z_{2})\op{c}{s_{1}}(z_{2})\op{c}{s_{2}}(z_{1})\ket{{\rm out}^{(2)}_{\{\xi\}}}}
{\bra{{\rm out}^{(2)}_{\{\xi\}}} \opd{c}{s_{1}}(z_{1})\op{c}{s_{1}}(z_{1})\ket{{\rm out}^{(2)}_{\{\xi\}}} \bra{{\rm out}^{(2)}_{\{\xi\}}} \opd{c}{s_{2}}(z_{2})\op{c}{s_{2}}(z_{2})\ket{{\rm out}^{(2)}_{\{\xi\}}}}$
where $(s_{1},s_{2} ) \in \{ R,L \} $. 
Here, we focus on the transmitted light, whose correlation function can be written as 
\begin{widetext}
\begin{eqnarray}
g_{RR}^{(2)} (z_{1},z_{2}) 
=\frac{
2 \Big\vert
\int_{\{ \xi(k)\}} \phi_{RR}(z_{1},z_{2})
\Big\vert^{2}
}
{
\frac{1}{M_{2}} 
\int dx  \Big(   \abs{\int_{\{ \xi(k)\}} \phi_{LR}(x,z_{1})}^{2}
+2 \abs{\int_{\{ \xi(k)\}} \phi_{RR}(x,z_{1})}^{2}
\Big)
\int dx  \Big(  \abs{\int_{\{ \xi(k)\}} \phi_{LR}(x,z_{2})}^{2}
+2 \abs{\int_{\{ \xi(k)\}} \phi_{RR}(x,z_{2})}^{2}
\Big)
}, 
\label{g2twophoton}
\end{eqnarray}
\end{widetext}
where $\int_{\{ \xi(k)\}} \equiv \int dk_{1}dk_{2} \xi_{1}(k_{1}) \xi_{2}(k_{2})$, and 
$\phi_{LR}^{(2)}$ and $\phi_{RR}^{(2)}$ represent the two-photon wave functions associated with Eqs.~(\ref{scatering_matrix_element_LR}), and (\ref{scatering_matrix_element_RR}), respectively (see Appendix~\ref{App_two_photon_scattering}). 
In this work, we will concentrate on the zero-delay case, i.e., $z_{1}=0$ and $z_{2}=0$. Note that the two-photon state of incoming light has different correlations for different values of $\Delta k$, since the distinguishability of the photons affects the intensity-intensity correlations.
Specifically, $g_{\rm initial}^{(2)}$ increases from $g_{\rm initial}^{(2)}=\frac{1}{2}$ when $k_{1}=k_{2}$ to $g_{\rm initial}^{(2)}=1$ when $\abs{k_{1}-k_{2}}\gg\sigma$ (see Appendix~\ref{App_initial_g2}). 

\begin{figure*}[t]
\centering
\includegraphics[width=18cm]{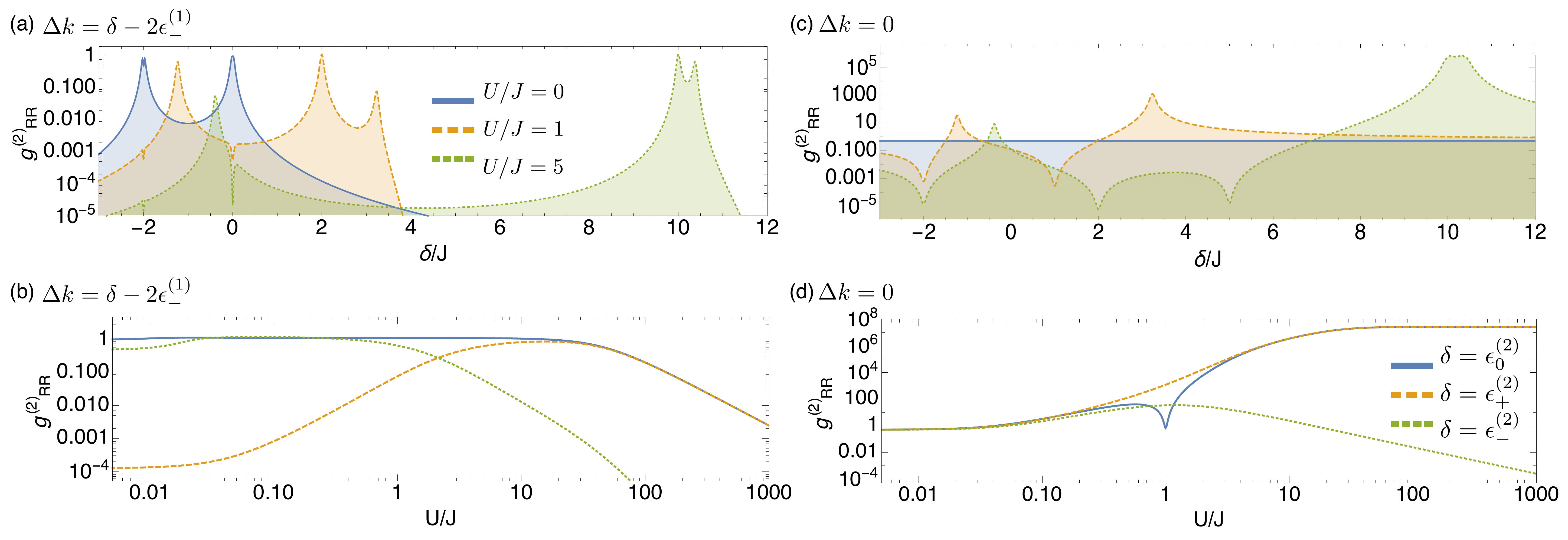}
\caption{Left- and right-hand columns : $\Delta k =\delta-2\epsilon_{-}^{(1)}$ and $\Delta k =0$. 
Second-order intensity correlation function, $g^{(2)}_{RR}$, is shown as a function of two-photon detuning $\delta/J$ for different photon-photon interaction strengths $U/J=0,1,5$ in (a) and (c). The correlation is also shown as a function of $U/J$ for two-photon eigenenergies
$\delta=\epsilon^{(2)}_{0,\pm}$ in (b) and (d).
The same parameters are chosen as used in Fig.~\ref{probabilities}.
We highlight here the direct mapping of the correlations of the many-body state $\epsilon_-^{(2)}$ onto the transmitted light $g^{(2)}_{RR}$ (the green dotted line in (b)), as the former approaches monotonically the Mott-like state $\ket{1,1}$ with increasing $U/J$. This does not hold in the identical-photons case however, where $g^{(2)}_{RR}$ first increases before it dips down to follow the correlations of the state.
 %large bunching is observed before $g^{(2)}_{RR}$ decreases and the latter dips below 1 only when $U/J > 20$. 
 The same behaviour is also found in the coherent-driving scenario \cite{Grujic13}. (See the detailed discussion in Section III. B regarding the rest of the states and regimes, and differences between the two approaches.)
}
\label{g2function} 
\end{figure*}

Figures~\ref{g2function}(a) and (c) plot the zero-delay second-order correlations against the total energy $\delta/J$. In the absence of nonlinearity, the $\Delta k =0$ case yields $g_{RR}^{(2)}=1/2$: being linear, the system does not change the statistics of the (identical) input photons. On the other hand, in the fully-resonant case, there are peaks when the photons have the energies $(\epsilon_-^{(1)},\epsilon_-^{(1)})$ and $(\epsilon_-^{(1)},\epsilon_+^{(1)})$, resulting in $g_{RR}^{(2)}\approx 1/2$ and $g_{RR}^{(2)}\approx 1$, respectively. Away from these points, $g^{(2)}_{RR} \sim 0$ because only one of the photons is transmitted. 
As the nonlinearity is introduced ($U=1,5$), correlations around the multi-photon correlated states change. Before we take a close look at these, there is an interesting observation worth describing; an anti-bunching observed at $\epsilon_{0}^{(2)}/2$ when $\Delta k=0$. This behaviour is not associated with any multi-photon correlated state, but arises due to a quantum interference between different path ways to the two-photon excitation in the second cavity. 
%Second, the intensity-intensity correlations for $\Delta k=0$ presented in Fig.~\ref{g2function} (c) is proportional to that obtained in the conventional coherent-driving scenario treated in the master equation formalism. In Appendix~\ref{App_masterequation} we compare the following three scenarios: 1) the coherent-field scattering, 2) the identical-photon scattering, and 3) the mater equation calculation with the semiclassical coherent driving term. We show that 1) and 3) produce practically the same $g^{(2)}$s, while 2) produces the $g^{(2)}$ that's a factor of two smaller--the factor that accounts for the initial state difference.

To see in detail how the second-order intensity correlations change with the interaction strength, we plot $g^{(2)}_{RR}$ as a function of $U/J$ at two-photon energies $\delta=\epsilon_{0,\pm}^{(2)}$ for the cases of $\Delta k =\delta - 2\epsilon_{-}^{(1)}$ in (b) and $\Delta k=0$ in (d).
Immediately, we note that over a wide range of $U/J$, the transmitted light at two-photon eigenenergies are anti-bunched (bunched) when $\Delta k =\delta - 2\epsilon_{-}^{(1)}$ ($\Delta k =0$).
Looking more closely, we find that $g^{(2)}_{RR}$ in the fully-resonant case provides a more faithful characterisation of the underlying multi-photon correlated states. Perhaps this is best illustrated by the $\delta = \epsilon_0^{(2)}$ (blue solid) curves. This state is proportional to $\ket{2,0}-\ket{0,2}$ regardless of the value of $U/J$, and therefore has a constant $g^{(2)}_{RR}$. This is exactly what is observed in the fully-resonant case in contrast to the identical-photons case, as long as the state is resolved from the state at $\epsilon_+^{(2)}$ (i.e., below $U/J \sim 10$). Similarly the $g^{(2)}_{RR}$ at $\epsilon_-^{(2)}$ shows the expected monotonic behaviour in the fully resonant case, due to the increase in $\ket{1,1}$ component with increasing $U/J$. In the identical-photons case, large bunching is observed before $g^{(2)}_{RR}$ decreases and dips below 1 only when $U/J > 20$. Similar behaviour is also found in the coherent-driving scenario \cite{Grujic13}. 

We attribute the qualitative differences between the two cases to the presence or absence of the resonant single-photon transmission. In the fully-resonant case, this is guaranteed by default and moreover the single photon transmission probability is robust at $\approx 1$ throughout a large range of $U/J$. This provides a nice constant background against which the second-order correlations can be measured. Such a background field is absent when $\Delta k = 0$ and bunching is generally observed because of suppressed single-photon transmission paired with enhanced two-photon transmission.
Incidentally, the little dip in Fig.~\ref{g2function}(d) ($g^{(2)}_{RR}=1/2$, the same as the background correlation) at $U/J=1$ is due to a single photon state (at $\epsilon_+^{(1)}$) coming into resonance with $\epsilon_0^{(2)}/2$.

From the above findings, we conclude that the fully-resonant scattering scenario exhibits a more faithful characteristics of the underlying many-body QCA states compared to the identical-photon scattering scenario.
%%%%%%%%%%%%%%%%%%%%%%%%%%%%%%%%%%%%%%%%%%%%%%%%%%%%%%%%%%%%%%%%%%%%%
\subsection{Comparison with the coherent-driving scenario}
%%%%%%%%%%%%%%%%%%%%%%%%%%%%%%%%%%%%%%%%%%%%%%%%%%%%%%%%%%%%%%%%%%%%%

\begin{figure}[b]
\centering
\includegraphics[width=8.5cm]{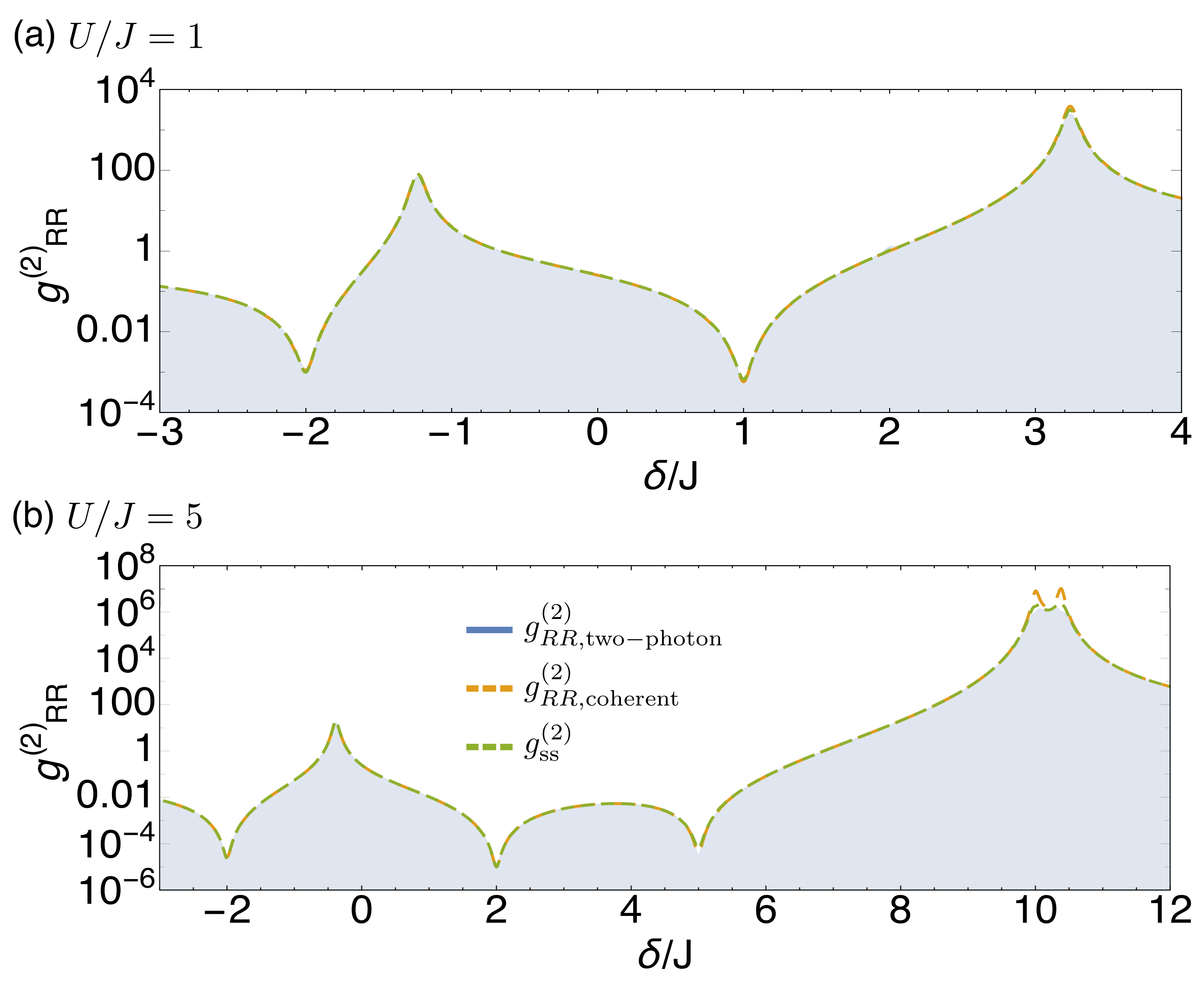}
\caption{
Second-order intensity correlations for the transmitted light as a function of the two-photon detuning $\delta/J$ for $U/J=1, 5$, obtained from the scattering approach with the identical two-photon input (blue solid) and a coherent-state input (orange dashed), and also from the master-equation formailsm (green dashed). For the latter, $\Omega=V\bar{n}=0.0002$ and $\gamma=V^{2}=0.04$ are used.  
}
\label{SMcomparison}
\end{figure}

Somewhat surprisingly, the intensity-intensity correlations for the identical-photons case are quantitatively very similar to those obtained from the coherent driving scenario. This can be seen by writing down the expressions for the correlation function in both cases.
%In this section, we compare the intensity-intensity correlations for $\Delta k=0$ presented in Figs.~\ref{g2function} (c) and (d) with the coherent-driving scenario. 
In the scattering formalism the coherent input field is incorporated by writing the input wave packet as $\ket{\alpha} = e^{\opd{c}{\alpha}-\op{c}{\alpha}}\ket{0}$ where $\opd{c}{\alpha}=\int dk \alpha(k) \hat{c}^{\dagger}(k)$ with the mean photon number $\bar{n}=\int dk \abs{\alpha(k)}^{2}$, and choose a Gaussian wave packet
\begin{eqnarray}
\alpha(k)=\frac{\sqrt{\bar{n}}}{(2\pi \sigma^{2})^{1/4}} {\rm exp} \Big( -\frac{(k-k_{c})^{2}}{4\sigma^{2}} \Big),\nonumber
\end{eqnarray}
where $\alpha(k)$ is narrowly peaked around $k_{c}$.
We here assume that the coherent-field is weak such that the mean photon number $\bar{n} \ll 1$. In this case, the output state $\ket{{\rm out}_{\alpha}}= \sum_{n}{\bf S}^{(n)} \ket{\alpha}$ can be approximated as
\begin{eqnarray}
\ket{{\rm out}_{\alpha}} \approx e^{-\bar{n}/2}( \ket{0} + {\bf S}^{(1)} \opd{c}{\alpha}\ket{0} + \frac{1}{2} {\bf S}^{(2)}(\opd{c}{\alpha})^{2} \ket{0}), \nonumber
\end{eqnarray}
where ${\bf S}^{(1)}$ and ${\bf S}^{(2)}$ are given as Eqs. (\ref{scatteringmatrix1}) and (\ref{scatteringmatrix2}), respectively. For the output state, the second-order intensity correlations can be calculated from 
\begin{eqnarray}
&& g_{RR,{\rm coherent}}^{(2)} (z_{1},z_{2}) \nonumber\\
&=& \frac{\bra{{\rm out}_{\alpha}} \opd{c}{R}(z_{1})\opd{c}{R}(z_{2})\op{c}{R}(z_{2})\op{c}{R}(z_{1})\ket{{\rm out}_{\alpha}}}
{\bra{{\rm out}_{\alpha}} \opd{c}{R}(z_{1})\op{c}{R}(z_{1})\ket{{\rm out}_{\alpha}}\bra{{\rm out}_{\alpha}} \opd{c}{R}(z_{2})\op{c}{R}(z_{2})\ket{{\rm out}_{\alpha}}} \label{g2coherent}\\
&\approx& \frac{1}{2} \frac{\vert \int_{\{ \alpha(k) \}} \phi_{RR}(z_{1},z_{2})\vert^{2}}{e^{-\bar{n}}\vert \int dk \alpha(k) \phi_{R}(z_{1}) \vert^{2} \vert \int dk \alpha(k) \phi_{R}(z_{2}) \vert^{2}}, \nonumber
\end{eqnarray} 
where $\int_{\{ \alpha(k)\}} \equiv \int dk_{1}dk_{2}\alpha(k_{1})\alpha(k_{2})$, and $\phi_{R}(x)$ represents the single-photon wave function (see Appendix~\ref{App_single_photon_scattering}).
On the other hand, the correlations in Eq.~(\ref{g2twophoton}) can be approximated for the identical two-photon input ($\Delta k = 0$) as 
\begin{eqnarray}
&& g_{RR,{\rm two-photon}}^{(2)}(z_{1},z_{2}) \nonumber\\
&\approx& \frac{1}{4} \frac{\vert \int_{\{ \xi(k) \}} \phi_{RR}(z_{1},z_{2})\vert^{2}}{\vert \int dk \xi(k) \phi_{R}(z_{1}) \vert^{2} \vert \int dk \xi(k) \phi_{R}(z_{2}) \vert^{2}}. \nonumber
\end{eqnarray}
One easily finds that the two cases only differ by a factor of $1/2$, identical to the difference in the initial correlations, i.e., 
\begin{eqnarray}
g_{RR,{\rm two-photon}}^{(2)} \approx \frac{1}{2} g_{RR,{\rm coherent}}^{(2)}. \label{twophoton_vs_coherent}
\end{eqnarray}
This is numerically demonstrated in Fig.~\ref{SMcomparison}, where we plot the zero-delay correlations in Eqs.~(\ref{g2twophoton}) (multiplied by $2$) and (\ref{g2coherent}) as a function of two-photon detuning $\delta/J$ for $U/J=1$ and $U/J=5$ when $V^{2}=0.04$ and $\bar{n}=0.001$. 
%Corrections come from the bound-state term $S_{RR}$, which can be neglected except near the resonances at $\epsilon_\pm^{(2)}$. 
In Fig.~\ref{SMcomparison}, we also compare with the conventional coherent-driving scenario treated in the master equation formalism, where the semiclassical coherent driving term $\Omega(\op{a}{1}+\opd{a}{1})$ is added in the $\hat{H}_{\rm cc}$ without considering $\hat{H}_{\rm wg}$ and $\hat{H}_{\rm wc}$, and then the second-order intensity correlations $g^{(2)}_{\rm ss}$ is calculated for the steady state $\rho_{\rm ss}$ obtained from a quantum optical master equation with a dissipation rate of $\gamma=V^{2}$. The numerical calculations of the master equation formalism show the consistent results as compared to the scattering approach for the coherent-state input, i.e., $g_{\rm ss}^{(2)}\approx g_{RR,{\rm coherent}}^{(2)}$.

%From the results in Fig.~\ref{SMcomparison} showing that the identical-photon scattering scenario and the conventional coherent-driving scenario exhibit the same behaviours as given in Eq.~(\ref{twophoton_vs_coherent}), 
From these results, we conclude that the fully-resonant scattering scenario has advantages over the conventional coherent-driving scenario in characterising the correlations of the underlying many-body QCA states.

\begin{figure}[b]
\centering
\includegraphics[width=8.5cm]{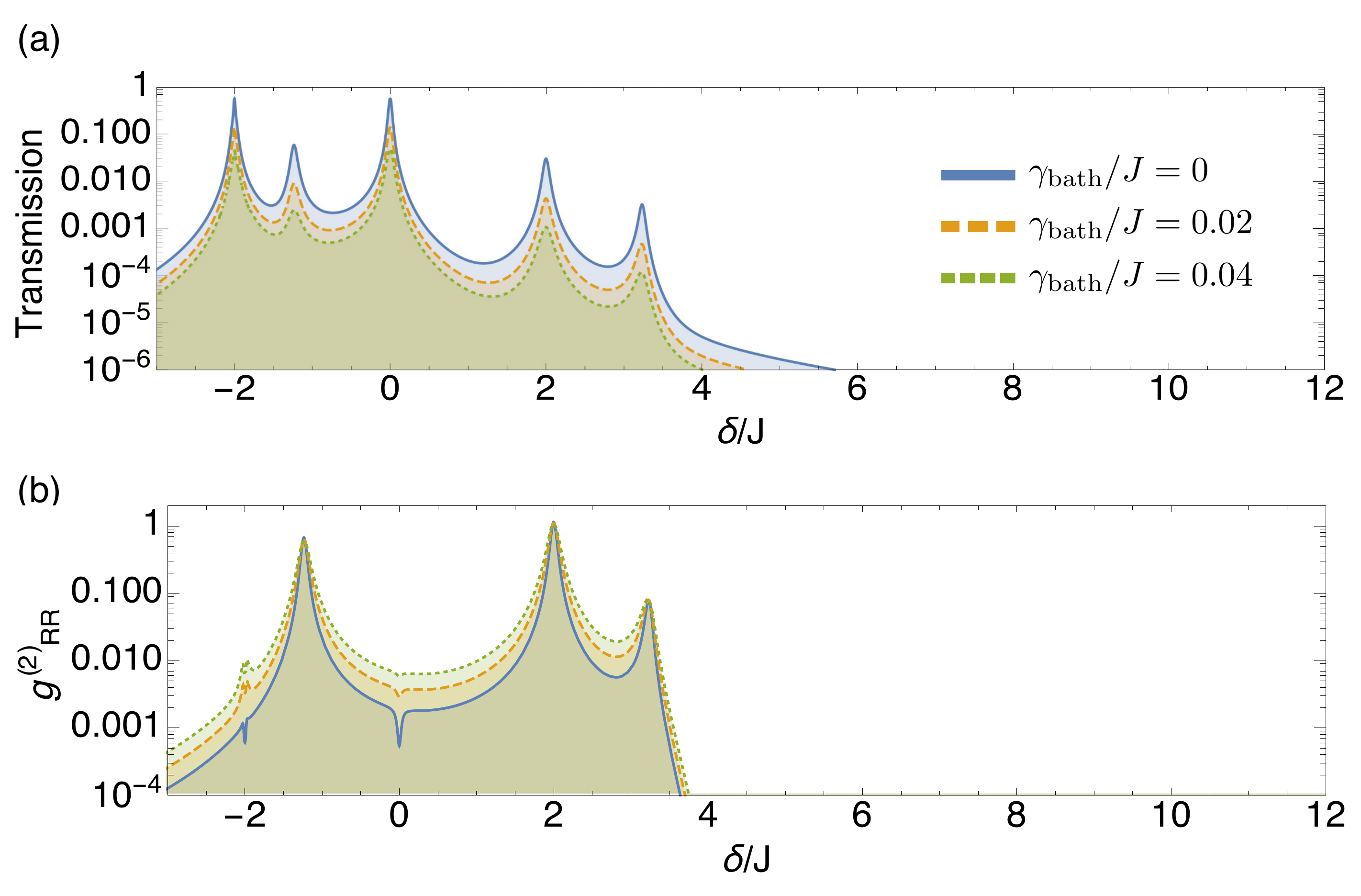}
\caption{For the fully-resonant case, two-photon transmissions ($P_{RR}$) and second-order correlations as a function of the total energy $\delta/J$ with $\gamma_{\rm bath}/J=0, 0.02$, and $0.04$ when $V^{2}/J=0.04$ and $U/J=1$. 
}
\label{losseffect}
\end{figure}
%%%%%%%%%%%%%%%%%%%%%%%%%%%%%%%%%%%%%%%%%%%%%%%%%%%%%%%%%%%%%%%%%%%%%
%%%%%%%%%%%%%%%%%%%%%%%%%%%%%%%%%%%%%%%%%%%%%%%%%%%%%%%%%%%%%%%%%%%%%
\section{Effects of photon losses}
Lastly, we address the issue of dissipation into non-guided modes. 
Within the scattering formalism used in this work, Markovian photon losses with the rate $\gamma_{\rm bath}$ can be accounted for by either introducing a waveguide for each cavity  \cite{Rephaeli13}, or equivalently using a combination of the scattering theory and the input-output formalism \cite{Fan10,Neumeier13}. 
In calculating the two-photon scattering matrices, it has been found that the effects of losses can be treated {\it exactly} by replacing the cavity frequency $\omega_{j}$ with $\omega_{j}-i\gamma_{\rm bath}/2$ in the Hamiltonian $\hat{H}_{\rm cc}$ \cite{Rephaeli13}. Using this method, we have calculated the two-photon transmission probability ($P_{RR}$) in the presence of extra photon losses in the cavities, as shown in Fig.~\ref{losseffect}(a). As expected, the transmission probability decreases and broadens as $\gamma_{\rm bath}$ increases while $V^2/J$ remains fixed. Things are a little more complicated for the second-order intensity correlation function $g^{(2)}_{RR}$. We must add extra contributions--in which one photon is in one of the extra loss channels--to the denominator of Eq.~(\ref{g2twophoton}). However, as a first consideration, one can ignore the effects of `quantum jumps' on these terms and calculate $g^{(2)}_{RR}$ using the non-Hermitian Hamiltonian. The results are plotted in Fig.~\ref{losseffect}(b), showing the effect of losses for $\gamma_{\rm bath}/J=0,0.02$, and $0.04$ when $V^{2}/J=0.04$ and $U/J=1$.

%%%%%%%%%%%%%%%%%%%%%%%%%%%%%%%%%%%%%%%%%%%%%%%%%%%%%%%%%%%%%%%%%%%%%
%%%%%%%%%%%%%%%%%%%%%%%%%%%%%%%%%%%%%%%%%%%%%%%%%%%%%%%%%%%%%%%%%%%%%
\section{Summary and Discussion}

To summarise, we have proposed a few-photon transport scenario to probe the many body structure of strongly correlated models simulated in QCAs.
We have demonstrated the feasibility of our proposal by analytically calculating the scattering matrix of the two-photon, two-site Bose-Hubbard QCA and studying the scattering probabilities and correlation functions. Signatures of strongly correlated multi-particle states were found in scattering probabilities and the second-order intensity correlations.  
We have compared two cases: 1) the fully-resonant case in which two input photons have tailored energies to match the single-particle and two-particle eigenenergies of the  model in question; 2) the identical-photons case in which two input photons have identical energies and are two-photon resonant with one of the two-particle states. We find that the multi-photon fully-resonant excitation scenario is advantageous over the alternative, in that it allows higher transmission probabilities and a more faithful mapping of the intensity-intensity correlations. Finally, we noted a correspondence between the identical-photon scattering case and the coherent-driving case, illustrating that the fully-resonant Fock-state scattering method has advantages over the latter.  The effects of losses in the cavities, as well as the incoming photons' pulse shapes and initial correlations are studied and analyzed.
%Finally, we showed that the identical-photons case displays \textcolor{blue}{the same} second-order correlations as the coherently-driven case, illustrating that the two-photon scattering method provides advantages over the latter.

A generalisation to larger arrays or number of photons is straightforward but the calculation is involved. To this end, field theoretic methods such as LSZ reduction formula \cite{Lehmann55}, or a general connection between the scattering matrix and Green's functions of the local system \cite{Xu15} might prove helpful in deducing the properties of higher $N$-photon scattering matrices which provides an interesting avenue for future research. Another interesting topic is to see whether a multi-coloured coherent driving fields can be used to obtain similar physics as studied in this work. We also note that our results are general and can be applied to probing the structure of any many-body bosonic models amenable to a QCA implementation including the Jaynes-Cummings-Hubbard, the extended Bose-Hubbard and a whole range of spin models.

Finally, we note that the scheme presented in this work can be experimentally demonstrated in a variety of systems, such as 
semiconductor microcavities \cite{Abbarchi13}, photonic crystal coupled cavities \cite{Majumdar12}, coupled optical waveguides \cite{Lepert11,Lepert13},
and superconducting circuits \cite{Houck12, Raftery14, Eichler14}. In the latter, a dimer array similar to the one we have described here has been fabricated and measured with high efficiency \cite{Raftery14, Eichler14}.
 
%%%%%%%%%%%%%%%%%%%%%%%%%%%%%%%%%%%%%%%%%%%%%%%%%%%%%%%%%%%%%%%%%%%%%
%%%%%%%%%%%%%%%%%%%%%%%%%%%%%%%%%%%%%%%%%%%%%%%%%%%%%%%%%%%%%%%%%%%%%
\acknowledgements
%%%%%%%%%%%%%%%%%%%%%%%%%%%%%%%%%%%%%%%%%%%%%%%%%%%%%%%%%%%%%%%%%%%%%
%%%%%%%%%%%%%%%%%%%%%%%%%%%%%%%%%%%%%%%%%%%%%%%%%%%%%%%%%%%%%%%%%%%%%
We thank D. E. Chang and M. Hartmann for helpful discussions, and C. Lee thanks P. N. Ma for useful comments about numerical calculations.
We would like to acknowledge the financial support provided by the National Research Foundation and Ministry of Education Singapore (partly through the Tier 3 Grant ``Random numbers from quantum processes''), and travel support by the EU IP-SIQS.

%%%%%%%%%%%%%%%%%%%%%%%%%%%%%%%%%%%%%%%%%%%%%%%%%%%%%%%%%%%%%%%%%%%%%
%%%%%%%%%%%%%%%%%%%%%%%%%%%%%%%%%%%%%%%%%%%%%%%%%%%%%%%%%%%%%%%%%%%%%
\appendix

%%%%%%%%%%%%%%%%%%%%%%%%%%%%%%%%%%%%%%%%%%%%%%%%%%%%%%%%%%%%%%%%%%%%%
%%%%%%%%%%%%%%%%%%%%%%%%%%%%%%%%%%%%%%%%%%%%%%%%%%%%%%%%%%%%%%%%%%%%%
\section{Scattering eigenstate}\label{App_scattering_eigenstate}
%%%%%%%%%%%%%%%%%%%%%%%%%%%%%%%%%%%%%%%%%%%%%%%%%%%%%%%%%%%%%%%%%%%%%
%%%%%%%%%%%%%%%%%%%%%%%%%%%%%%%%%%%%%%%%%%%%%%%%%%%%%%%%%%%%%%%%%%%%%
In this section, we provide a detailed derivation of the scattering matrices for the single- and two-photons cases.  
%%%%%%%%%%%%%%%%%%%%%%%%%%%%%%%%%%%%%%%%%%%%%%%%%%%%%%%%%%%%%%%%%%%%%
%%%%%%%%%%%%%%%%%%%%%%%%%%%%%%%%%%%%%%%%%%%%%%%%%%%%%%%%%%%%%%%%%%%%%
\subsection{Single-photon scattering}\label{App_single_photon_scattering}
%%%%%%%%%%%%%%%%%%%%%%%%%%%%%%%%%%%%%%%%%%%%%%%%%%%%%%%%%%%%%%%%%%%%%
%%%%%%%%%%%%%%%%%%%%%%%%%%%%%%%%%%%%%%%%%%%%%%%%%%%%%%%%%%%%%%%%%%%%%
Single-photon scattering eigenstates are written as 
\begin{eqnarray}
\ket{E^{(1)}}=\int_{-\infty}^{\infty}dx \phi_{L}(x)\opd{c}{L}(x)\ket{0}+\int_{-\infty}^{\infty}dy \phi_{R}(y)\opd{c}{R}(y)\ket{0} \nonumber\\
+e_{1}\opd{a}{1}\ket{0}+e_{2}\opd{a}{2}\ket{0}.
\end{eqnarray}
The time independent Schr\"odinger equation $\hat{H}_{\rm tot}\ket{E^{(1)}}=E^{(1)}\ket{E^{(1)}}$  with $E^{(1)}=k$ leads to the following set of equations,
\begin{eqnarray}
-i\frac{\partial}{\partial x} \phi_{L}(x) + V_{1}\delta(x)e_{1}&=&E^{(1)} \phi_{L}(x), \label{single1}\\
-i\frac{\partial}{\partial y} \phi_{R}(y) + V_{2}\delta(y)e_{2}&=&E^{(1)} \phi_{R}(y), \label{single2}\\
\omega_{1} e_{1}+Je_{2}+V_{1}\phi_{L}(0)&=&E^{(1)} e_{1}, \label{single3}\\
\omega_{2} e_{2}+Je_{1}+V_{2}\phi_{R}(0)&=&E^{(1)} e_{2}. \label{single4}
\end{eqnarray}
From Eqs.~(\ref{single1}) and (\ref{single2}), the discontinuity relations are given by $\phi_{L}(0_{+}) = \phi_{L}(0_{-}) -iV_{1}e_{1} = \frac{1}{\sqrt{2\pi}} -iV_{1}e_{1}$ and $\phi_{R}(0_{+}) =\phi_{R}(0_{-}) -iV_{2}e_{2} = -iV_{2}e_{2}$, provided that the initial regions are considered as $\phi_{L}(x<0) =\frac{1}{\sqrt{2\pi}} e^{ikx}$ and $\phi_{R}(y<0)=0$.
Furthermore, we have $\phi_{L}(0) = \frac{1}{2}(\phi_{L}(0_{+})+\phi_{L}(0_{-}))$ and $\phi_{R}(0) = \frac{1}{2}(\phi_{R}(0_{+})+\phi_{R}(0_{-}))$.
Now, solving Eqs.~(\ref{single1}) and (\ref{single2}) in the region $x>0$ and $y>0$ one finds 
\begin{eqnarray}
\phi_{L}(x) &=& \frac{1}{\sqrt{2\pi}}(\theta(-x)+r_{k}\theta(x))e^{ikx}\nonumber\\
\phi_{R}(y) &=& \frac{1}{\sqrt{2\pi}} t_{k}\theta(y)  e^{iky}.\nonumber
\end{eqnarray}
The transmission and reflection coefficients are found from the relations $r_{k} =-\sqrt{2\pi}ie_{1}V_{1} +1$ and $t_{k} = -\sqrt{2\pi}ie_{2}V_{2} $, where $e_{1}$ and $e_{2}$ are calculated from Eqs.~(\ref{single3}) and (\ref{single4}):
\begin{eqnarray}
e_{1} &=& \frac{\sqrt{\frac{2}{\pi}} V_{1} (-i V_{2}^{2}-2E^{(1)}+2\omega_{2})}{4J^{2}+(V_{1}^{2}-2i(E^{(1)}-\omega_{1}))(V_{2}^{2}-2i(E^{(1)}-\omega_{2}))}, \nonumber\\
e_{2} &=& -\frac{2J\sqrt{\frac{2}{\pi}}V_{1}}{4J^{2}+(V_{1}^{2}-2i(E^{(1)}-\omega_{1}))(V_{2}^{2}-2i(E^{(1)}-\omega_{2}))}. \nonumber
\end{eqnarray}
Thus, the explicit expressions of transmission and reflection coefficients are written as
\begin{eqnarray}
r_{k}&=&\frac{4J^{2}-(V_{1}^{2}+2i(E^{(1)}-\omega_{1}))(V_{2}^{2}-2i(E^{(1)}-\omega_{2}))}{4J^{2}+(V_{1}^{2}-2i(E^{(1)}-\omega_{1}))(V_{2}^{2}-2i(E^{(1)}-\omega_{2}))}, \nonumber\\
&&\label{rk}\\
t_{k}&=&\frac{4iJV_{1}V_{2}}{4J^{2}+(V_{1}^{2}-2i(E^{(1)}-\omega_{1}))(V_{2}^{2}-2i(E^{(1)}-\omega_{2}))}.\nonumber\\
&&\label{tk}
\end{eqnarray}
  As expected, nonlinear effects do not appear in this single photon case, and hence the transmission and reflection of single photons are equivalent to the case of two two-level atoms \cite{Roy13}, or two linear resonators \cite{Lee12}.
Using these results, $\phi_{L}(x)$ and $\phi_{R}(y)$ construct the single photon scattering matrix \cite{Zheng10, Shen07a}
\begin{eqnarray}
{\bf S}^{(1)}=\int dk \ket{\phi_{\rm out}^{(1)}}_{k}\bra{\phi_{\rm in}^{(1)}},
\label{scatteringmatrix1}
\end{eqnarray} 
where the input and output states of single photon are written as $\ket{\phi_{\rm in}^{(1)}}_{k} = \int dx \phi_{L}(x{\rm<}0) \opd{c}{L}(x) \ket{0}$ and $\ket{\phi_{\rm out}^{(1)}}_{k} = \ket{\phi_{\rm out}^{(1)}}_{L} + \ket{\phi_{\rm out}^{(1)}}_{R}$, with $\ket{\phi_{\rm out}^{(1)}}_{L} = \int dx \phi_{L}(x{\rm>}0) \opd{c}{L}(x) \ket{0} $ and $\ket{\phi_{\rm out}^{(1)}}_{R} = \int dy \phi_{R}(y{\rm>}0) \opd{c}{R}(y) \ket{0} $.

%%%%%%%%%%%%%%%%%%%%%%%%%%%%%%%%%%%%%%%%%%%%%%%%%%%%%%%%%%%%%%%%%%%%%
%%%%%%%%%%%%%%%%%%%%%%%%%%%%%%%%%%%%%%%%%%%%%%%%%%%%%%%%%%%%%%%%%%%%%
\subsection{Two-photon scattering}\label{App_two_photon_scattering}
%%%%%%%%%%%%%%%%%%%%%%%%%%%%%%%%%%%%%%%%%%%%%%%%%%%%%%%%%%%%%%%%%%%%%
%%%%%%%%%%%%%%%%%%%%%%%%%%%%%%%%%%%%%%%%%%%%%%%%%%%%%%%%%%%%%%%%%%%%%
For the two-photon scattering problem, a general form of two-photon eigenstates $\ket{E^{(2)}}=\ket{E_{1}^{(2)}}+\ket{E_{2}^{(2)}}+\ket{E_{3}^{(2)}}$ is given as
\begin{eqnarray}
&&\ket{E_{1}^{(2)}}=\int_{-\infty}^{\infty} dx_{1}dx_{2} \phi_{LL}(x_{1},x_{2}) \frac{1}{\sqrt{2}} \opd{c}{L}(x_{1}) \opd{c}{L}(x_{2}) \ket{0} \nonumber\\
&&~~~~~~~~~~~+\int_{-\infty}^{\infty} dy_{1}dy_{2} \phi_{RR}(y_{1},y_{2}) \frac{1}{\sqrt{2}}\opd{c}{R}(y_{1}) \opd{c}{R}(y_{2}) \ket{0} \nonumber\\
&&~~~~~~~~~~~~~~~+\int_{-\infty}^{\infty} dx_{1}dy_{1} \phi_{LR}(x_{1},y_{1}) \opd{c}{L}(x_{1}) \opd{c}{R}(y_{1}) \ket{0}, \nonumber\\
&&\ket{E_{2}^{(2)}}= e_{11} \frac{1}{\sqrt{2}}\opd{a}{1}\opd{a}{1} \ket{0} + e_{12}\opd{a}{1}\opd{a}{2}\ket{0}+e_{22}\frac{1}{\sqrt{2}}\opd{a}{2}\opd{a}{2}\ket{0}, \nonumber\\
&&\ket{E_{3}^{(2)}}= \int_{-\infty}^{\infty} dx_{1} \Big( \phi_{L1}(x_{1}) \opd{c}{L}(x_{1})\opd{a}{1} +\phi_{L2}(x_{1}) \opd{c}{L}(x_{1})\opd{a}{2}  \Big) \ket{0} \nonumber\\
&&~~~~~~~~~~~+\int_{-\infty}^{\infty} dy_{1} \Big( \phi_{R1}(y_{1}) \opd{c}{R}(y_{1})\opd{a}{1} +\phi_{R2}(y_{1}) \opd{c}{R}(y_{1})\opd{a}{2}  \Big) \ket{0}.\nonumber
\end{eqnarray}
$\ket{E_{1}^{(2)}}$ represents two photons in either the left or right waveguide, $\ket{E_{2}^{(2)}}$ represents two photons in the coupled cavities, and $\ket{E_{3}^{(2)}}$ describes one photon in one of the waveguides and the other in one of the cavities. We here obtain the two-photon scattering eigenstates by imposing the open boundary condition.
The Schr\"odinger equation $\hat{H}_{\rm tot}\ket{E^{(2)}}=E^{(2)}\ket{E^{(2)}}$ gives
\begin{widetext}
\begin{eqnarray}
 -i \frac{\partial}{\partial x_{2}}\phi_{LL}(x_{1},x_{2}) -i \frac{\partial}{\partial x_{1}}\phi_{LL}(x_{1},x_{2}) +\frac{V_{1}}{\sqrt{2}} ( \delta(x_{1}) \phi_{L1}(x_{2}) + \delta(x_{2}) \phi_{L1}(x_{1}))  &=& E^{(2)} \phi_{LL}(x_{1},x_{2}), \label{two1}\\
 -i \frac{\partial}{\partial x_{1}}\phi_{LR}(x_{1},y_{1}) -i \frac{\partial}{\partial y_{1}}\phi_{LR}(x_{1},y_{1}) + V_{1}\delta(x_{1}) \phi_{R1}(y_{1}) + V_{2}\delta(y_{1}) \phi_{L2}(x_{1})  &=&E^{(2)} \phi_{LR}(x_{1},y_{1}), \label{two2}\\
 -i \frac{\partial}{\partial y_{2}}\phi_{RR}(y_{1},y_{2}) -i \frac{\partial}{\partial y_{1}}\phi_{RR}(y_{1},y_{2}) + \frac{V_{2}}{\sqrt{2}}(\delta(y_{1}) \phi_{R2}(y_{2})+ \delta(y_{2}) \phi_{R2}(y_{1})) &=&E^{(2)} \phi_{RR}(y_{1},y_{2}),\label{two3}\\
\nonumber\\
-i\frac{\partial}{\partial x_{1}} \phi_{L1}(x_{1}) + \phi_{L1}(x_{1})\omega_{1}+\phi_{L2}(x_{1})J + V_{1}\delta(x_{1})e_{11}\sqrt{2} + V_{1}\frac{1}{\sqrt{2}}(\phi_{LL}(x_{1},0)+\phi_{LL}(0,x_{1})) &=&E^{(2)} \phi_{L1}(x_{1}), \label{two4}\\
-i\frac{\partial}{\partial x_{1}} \phi_{L2}(x_{1}) + \phi_{L2}(x_{1})\omega_{2}+\phi_{L1}(x_{1})J + V_{1}\delta(x_{1})e_{12} +V_{2} \phi_{LR}(x_{1},0) &=&E^{(2)} \phi_{L2}(x_{1}), \label{two5}\\
-i\frac{\partial}{\partial y_{1}} \phi_{R1}(y_{1}) + \phi_{R1}(y_{1})\omega_{1}+\phi_{R2}(y_{1})J + V_{2}\delta(y_{1})e_{12} +V_{1}\phi_{LR}(0,y_{1}) &=&E^{(2)} \phi_{R1}(y_{1}), \label{two6}\\
-i\frac{\partial}{\partial y_{1}} \phi_{R2}(y_{1}) + \phi_{R2}(y_{1})\omega_{2}+\phi_{R1}(y_{1})J + V_{2}\delta(y_{1})e_{22}\sqrt{2} +V_{2} \frac{1}{\sqrt{2}}(\phi_{RR}(y_{1},0)+\phi_{RR}(0,y_{1})) &=&E^{(2)} \phi_{R2}(y_{1}), \label{two7}\\
\nonumber\\
\omega_{1}e_{11}\sqrt{2}+Je_{12}+U_{1}e_{11}\sqrt{2} + V_{1}\phi_{L1}(0) &=&E^{(2)} e_{11} \frac{1}{\sqrt{2}},\label{two8}\\
\omega_{1}e_{12}+\omega_{2}e_{12}+Je_{22}\sqrt{2}+Je_{11}\sqrt{2} +V_{1}\phi_{L2}(0)+V_{2}\phi_{R1}(0) &=&E^{(2)} e_{12}, \label{two9}\\
\omega_{2}e_{22}\sqrt{2}+Je_{12}+U_{2}e_{22}\sqrt{2} + V_{2}\phi_{R2}(0) &=&E^{(2)} e_{22}\frac{1}{\sqrt{2}}. \label{two10}
\end{eqnarray}
\end{widetext}
Let us first solve these equations in the half spaces, $x_{1}<x_{2}$, $x_{1}<y_{1}$, and $y_{1}<y_{2}$.
In this case, there are three quadrants: \ding{192} $x_{1}<x_{2}<0$, $x_{1}<y_{1}<0$, $y_{1}<y_{2}<0$, \ding{193} $x_{1}<0<x_{2}$, $x_{1}<0<y_{1}$, $y_{1}<0<y_{2}$, and \ding{194} $0<x_{1}<x_{2}$, $0<x_{1}<y_{1}$, $0<y_{1}<y_{2}$. The Initial conditions for the amplitudes in the region \ding{192} are given as
\begin{eqnarray}
\phi_{LL}(x_{1}<0,x_{2}<0) &=& \frac{1}{\sqrt{2}}\frac{1}{2\pi}(e^{ik_{1}x_{1}+ik_{2}x_{2}}+e^{ik_{2}x_{1}+ik_{1}x_{2}}), \nonumber\\
&& \label{twoinitial1}\\
\phi_{LR}(x_{1}<0,y_{1}<0) &=& 0, \label{twoinitial2}\\
\phi_{RR}(y_{1}<0,y_{2}<0) &=& 0 . \label{twoinitial3}
\end{eqnarray}
The discontinuity relations of the two-photon amplitudes across $x_{1}, x_{2}, y_{1}, y_{2}=0$ are given from Eqs.~(\ref{two1})-(\ref{two3}):
\begin{eqnarray}
\phi_{LL}(0_{+},x_{2})&=&\phi_{LL}(0_{-},x_{2})-i\frac{V_{1}}{\sqrt{2}}\phi_{L1}(x_{2}), \label{twodis1}\\
\phi_{LL}(x_{1},0_{+})&=&\phi_{LL}(x_{1},0_{-})-i\frac{V_{1}}{\sqrt{2}}\phi_{L1}(x_{1}), \label{twodis2}\\
\phi_{LR}(0_{+},y_{1})&=&\phi_{LR}(0_{-},y_{1})-iV_{1}\phi_{R1}(y_{1}), \label{twodis3}\\
\phi_{LR}(x_{1},0_{+})&=&\phi_{LR}(x_{1},0_{-})-iV_{2}\phi_{L2}(x_{1}), \label{twodis4}\\
\phi_{RR}(0_{+},y_{2})&=&\phi_{RR}(0_{-},y_{2})-i\frac{V_{2}}{\sqrt{2}}\phi_{R2}(y_{2}), \label{twodis5}\\
\phi_{RR}(y_{1},0_{+})&=&\phi_{RR}(y_{1},0_{-})-i\frac{V_{2}}{\sqrt{2}}\phi_{R2}(y_{1}). \label{twodis6}
\end{eqnarray}
Similarly, the discontinuity relations of the cavity-photon amplitudes across the origin are given from Eqs.~(\ref{two4})-(\ref{two7}):
\begin{eqnarray}
\phi_{L1}(0_{+})&=&\phi_{L1}(0_{-})-iV_{1}e_{11}\sqrt{2}, \label{twodis7}\\
\phi_{L2}(0_{+})&=&\phi_{L2}(0_{-})-iV_{1}e_{12}, \label{twodis8}\\
\phi_{R1}(0_{+})&=&\phi_{R1}(0_{-})-iV_{2}e_{12}, \label{twodis9}\\
\phi_{R2}(0_{+})&=&\phi_{R2}(0_{-})-iV_{2}e_{22}\sqrt{2}. \label{twodis10}
\end{eqnarray}
Two-photon and cavity-photon amplitudes are also discontinuous at $x_{1}, x_{2}, y_{1}, y_{2}=0$ and therefore we set
\begin{eqnarray}
\phi_{LL}(0,x)&=&\phi_{LL}(x,0)=\frac{1}{2}(\phi_{LL}(0_{+},x)+\phi_{LL}(0_{-},x)),\nonumber\\
&& \label{tworeg1}\\
\phi_{LR}(0,y)&=&\frac{1}{2}(\phi_{LR}(0_{+},y)+\phi_{LR}(0_{-},y)), \label{tworeg2}\\
\phi_{LR}(x,0)&=&\frac{1}{2}(\phi_{LR}(x,0_{+})+\phi_{LR}(x,0_{-})), \label{tworeg3}\\
\phi_{RR}(0,y)&=&\phi_{RR}(y,0)=\frac{1}{2}(\phi_{RR}(0_{+},y)+\phi_{RR}(0_{-},y)).\nonumber\\
\label{tworeg4}
\end{eqnarray}

From these, the coupled linear inhomogeneous first-order differential equations (\ref{two4}), (\ref{two5}), (\ref{two6}), and (\ref{two7}) in region \ding{192} can be rewritten as
\begin{widetext}
\begin{eqnarray}
i \frac{\partial}{\partial x} 
\begin{pmatrix}
\phi_{L1}(x<0) \\ \phi_{L2}(x<0)
\end{pmatrix}
&=&
\begin{pmatrix}
\omega_{1}-E^{(2)}-\frac{iV_{1}^{2}}{2} & J \\
J&\omega_{2}-E^{(2)}-\frac{iV_{2}^{2}}{2}
\end{pmatrix}
\begin{pmatrix}
\phi_{L1}(x<0) \\ \phi_{L2}(x<0)
\end{pmatrix}
+
\begin{pmatrix}
\sqrt{2}V_{1}\phi_{LL}(x<0,0_{-}) \\ V_{2} \phi_{LR}(x<0,0_{-})
\end{pmatrix}
+
\begin{pmatrix}
V_{1}e_{11}\sqrt{2} \\ V_{1}e_{12}
\end{pmatrix}
\delta(x) \label{deq1}, \\
%%%%%%%%%%%%%%%%%%%%%%
i \frac{\partial}{\partial y} 
\begin{pmatrix}
\phi_{R1}(y<0) \\ \phi_{R2}(y<0)
\end{pmatrix}
&=&
\begin{pmatrix}
\omega_{1}-E^{(2)}-\frac{iV_{1}^{2}}{2} & J \\
J&\omega_{2}-E^{(2)}-\frac{iV_{2}^{2}}{2}
\end{pmatrix}
\begin{pmatrix}
\phi_{R1}(y<0) \\ \phi_{R2}(y<0)
\end{pmatrix}
+
\begin{pmatrix}
V_{1}\phi_{LR}(0_{-},y<0) \\ \sqrt{2}V_{2} \phi_{RR}(0_{-},y<0)
\end{pmatrix}
+
\begin{pmatrix}
V_{2}e_{12} \\ V_{2}e_{22}\sqrt{2}
\end{pmatrix}
\delta(y) . \label{deq2}
\end{eqnarray}
\end{widetext}
We solve these with the discontinuity relations and the initial conditions in Eq.~(\ref{twoinitial1})-(\ref{twoinitial3}) to find 
\begin{eqnarray}
\phi_{L1}(x<0) &=&\frac{1}{\sqrt{2}}\frac{1}{2\pi} ( \chi_{L1k_{2}}e^{ik_{1}x}+\chi_{L1k_{1}}e^{ik_{2}x}), \label{phil1m}\\ 
\phi_{L2}(x<0) &=&\frac{1}{\sqrt{2}}\frac{1}{2\pi} ( \chi_{L2k_{2}}e^{ik_{1}x}+\chi_{L2k_{1}} e^{ik_{2}x}), \label{phil2m}\\
\phi_{R1}(y<0) &=& 0, \label{phir1m}\\
\phi_{R2}(y<0)&=& 0, \label{phir2m}
\end{eqnarray}
where
\begin{eqnarray}
\chi_{L1k_{1}}&=&A\Big(\frac{M_{-}}{(k_{2}+\lambda_{-})}-\frac{M_{+}}{(k_{2}+\lambda_{+})} \Big), \nonumber\\
\chi_{L1k_{2}}&=&A\Big( \frac{M_{-}}{(k_{1}+\lambda_{-})} -\frac{M_{+}}{(k_{1}+\lambda_{+})} \Big), \nonumber\\
\chi_{L2k_{1}}&=&A\Big( \frac{1}{(k_{2}+\lambda_{-})}-\frac{1}{(k_{2}+\lambda_{+})} \Big), \nonumber\\
\chi_{L2k_{2}}&=&A\Big( \frac{1}{(k_{1}+\lambda_{-})}-\frac{1}{(k_{1}+\lambda_{+})} \Big), \nonumber\\
\nonumber\\
A&=&\frac{2\sqrt{2}V_{1}J}{\sqrt{16J^{2}-(V_{1}^{2}-V_{2}^{2}+2i(\omega_{1}-\omega_{2}))^{2}}},  \nonumber\\
M_{\mp}&=& \frac{-i V_{1}^{2} + i V_{2}^{2} + 2 (\omega_{1} - \omega_{2}) \mp 2\sqrt{2}V_{1}J/A}{4 J}, \nonumber\\
\lambda_{\mp}&=&\frac{1}{4}\Big( -iV_{1}^{2}-i V_{2}^{2}-4E^{(2)}+2(\omega_{1}+\omega_{2})\mp2\sqrt{2}V_{1}J/A \Big).\nonumber\\
\end{eqnarray}

%%%%%%%%%%%%%%%%%%%%%%%%%%%%%%%%%%%%%%%%%%%%%%%%%%%%%%%%%%%%%%%%%%%%%
Substituting Eqs.~(\ref{twodis7})-(\ref{twodis10}) into Eqs.~(\ref{two8})-(\ref{two10}), we obtain $e_{11}, e_{12}$ and $e_{22}$ as follows
\begin{widetext}
\begin{eqnarray}
e_{11}&=&\sqrt{2}V_{1}\Big(
4J^{2}\phi_{L1}(0_{-})+\phi_{L1}(0_{-})(V_{1}^{2}+V_{2}^{2}-2i(E^{(2)}-\omega_{1}-\omega_{2}))\nonumber\\
&&~~~~~~~~~~~~~~~~~~~~~\times (2iU_{2}+V_{2}^{2}-iE^{(2)}+2i\omega_{2})+2J\phi_{L2}(0_{-})(2U_{2}-iV_{2}^{2}-E^{(2)}+2\omega_{2}) \Big)/\eta, \label{e11}\\
e_{12}&=&2V_{1}\Big(2J\phi_{L1}(0_{-}) +\phi_{L2}(0_{-}) (-2U_{1}+iV_{1}^{2}+E^{(2)}-2\omega_{1})\Big)(-2U_{2}+iV_{2}^{2}+E^{(2)}-2\omega_{2}) /\eta, \label{e12}\\
e_{22}&=& 2\sqrt{2}JV_{1}\Big( 2J\phi_{L1}(0_{-}) +\phi_{L2}(0_{-})(-2U_{1}+iV_{1}^{2}+E^{(2)}-2\omega_{1})\Big) / \eta,\label{e22},\\
\eta&=&\Big((2iU_{1}+V_{1}^{2}-iE^{(2)}+2i\omega_{1}) (V_{1}^{2}+V_{2}^{2}-2i(E^{(2)}-\omega_{1}-\omega_{2}))(2U_{2}-iV_{2}^{2}-E^{(2)}+2\omega_{2})\nonumber\\
&&+J^{2}(8U_{1}+8U_{2}-4iV_{1}^{2}-4iV_{2}^{2}+8(-E^{(2)}+\omega_{1}+\omega_{2}))\Big).\nonumber
\end{eqnarray}
\end{widetext}

\noindent Here, we note that the amplitudes of two-photon excitations to be in the same cavity, $e_{11}$ and $e_{22}$, approach zero in the limit of $U_{1}$ and $U_{2} \rightarrow \infty$ as these two-photon excitations require an infinite amount of energy.

%%%%%%%%%%%%%%%%%%%%%%%%%%%%%%%%%%%%%%%%%%%%%%%%%%%%%%%%%%%%%%%%%%%%%
Substituting the initial conditions in region \ding{192} and Eqs.~(\ref{phil1m}), (\ref{phil2m}), (\ref{phir1m}), and (\ref{phir2m}) into the discontinuity relations, we obtain
\begin{eqnarray}
\phi_{LL}(x_{1}<0,0_{+})&=& \frac{1}{\sqrt{2}}\frac{1}{2\pi} (r_{k_{2}}e^{ik_{1}x_{1}}+r_{k_{1}}e^{ik_{2}x_{1}}) ,\label{phillm0}\\
\phi_{LR}(x_{1}<0,0_{+})&=& \frac{1}{2\pi} ( t_{k_{2}} e^{ik_{1}x_{1}}+ t_{k_{1}} e^{ik_{2}x_{1}}), \label{philrm0}\\
\phi_{RR}(y_{1}<0,0_{+})&=&=0, \label{phirrm0}
\end{eqnarray}
where the single-photon transmission and reflection coefficients for $E^{(2)}=k_{1}+k_{2}$ are defined as
\begin{eqnarray}
r_{k_{j}} &=& (1-i\frac{V_{1}}{\sqrt{2}}\chi_{L1k_{j}} ), \nonumber\\
t_{k_{j}} &=& -i\frac{V_{2}}{\sqrt{2}}\chi_{L2k_{j}},  \nonumber
\end{eqnarray}
where $j=1,2$. These are same as Eqs.~(\ref{rk}) and (\ref{tk}).
We now solve Eqs.~(\ref{two1})-(\ref{two3}) in region \ding{193} with the initial conditions in Eqs.~(\ref{phillm0}) - (\ref{phirrm0}) to find
\begin{eqnarray}
\phi_{LL}(x_{1}<0,x_{2}>0) &=&\frac{1}{\sqrt{2}}\frac{1}{2\pi}( r_{k_{2}} e^{ik_{1} x_{1}+ik_{2} x_{2}} +r_{k_{1}} e^{ik_{2}x_{1}+ik_{1} x_{2}} ), \nonumber\\ &&\label{phillmp}\\
\phi_{LR}(x_{1}<0,y_{1}>0) &=&\frac{1}{2\pi}( t_{k_{2}}e^{ik_{1}x_{1} + i k_{2} y_{1}} +t_{k_{1}}e^{ik_{2}x_{1} + i k_{1} y_{1}}),  \label{philrmp}\\
\phi_{RR}(y_{1}<0,y_{2}>0) &=& 0. \label{phirrmp}
\end{eqnarray}

%%%%%%%%%%%%%%%%%%%%%%%%%%%%%%%%%%%%%%%%%%%%%%%%%%%%%%%%%%%%%%%%%%%%%
Then solving eqs.~(\ref{two4}), (\ref{two5}), (\ref{two6}), and (\ref{two7}) in region \ding{194} with the boundary conditions for $\phi_{L1}(0_{+})$, $\phi_{L2}(0_{+})$, $\phi_{R1}(0_{+})$, $\phi_{R2}(0_{+})$, $\phi_{LL}(x>0,0_{-})$, $\phi_{LR}(x>0,0_{-})$, $\phi_{LR}(0_{-},y>0)$, $\phi_{RR}(0_{-},y>0)$, we obtain
\begin{eqnarray}
&\phi_{L1}(x>0) &= \frac{1}{\sqrt{2}} \frac{1}{2\pi}\Big( r_{k_{1}}\chi_{L1k_{2}}e^{i k_{1}x} + r_{k_{2}}\chi_{L1k_{1}} e^{i k_{2}x} \nonumber\\
&&~~~~+ M_{-} c_{L_{-}}e^{-i \lambda_{-} x} + M_{+}c_{L_{+}}e^{-i \lambda_{+} x}\Big), \label{phil1p}\\
&\phi_{L2}(x>0) &= \frac{1}{\sqrt{2}} \frac{1}{2\pi} \Big( r_{k_{1}}\chi_{L2k_{2}}e^{i k_{1}x} +r_{k_{2}}\chi_{L2k_{1}} e^{i k_{2}x} \nonumber\\
&&~~~~+ c_{L_{-}}e^{-i \lambda_{-} x}+c_{L_{+}}e^{-i \lambda_{+} x} \Big), \label{phil2p}\\
&\phi_{R1}(y>0)&= \frac{1}{\sqrt{2}} \frac{1}{2\pi} \Big( t_{k_{1}}\chi_{L1k_{2}}e^{i k_{1} y} + t_{k_{2}}\chi_{L1k_{1}}e^{ik_{2}y} \nonumber\\
&&~~~~+M_{-}c_{R_{-}} e^{-i\lambda_{-}y}+ M_{+}c_{R_{+}}e^{-i \lambda_{+}y} \Big), \label{phir1p}\\
&\phi_{R2}(y>0)&= \frac{1}{\sqrt{2}} \frac{1}{2\pi} \Big( t_{k_{1}}\chi_{L2k_{2}} e^{i k_{1} y} +t_{k_{2}}\chi_{L2k_{1}}e^{ik_{2}y} \nonumber\\
&&~~~~+c_{R_{-}} e^{-i\lambda_{-}y}+ c_{R_{+}}e^{-i \lambda_{+}y}\Big), \label{phir2p}
\end{eqnarray}
where
\begin{eqnarray}
c_{L_{\mp}} &=& \pm A\Big(\frac{2\pi}{V_{1}} \big( M_{\pm} \phi_{L2}(0_{+}) - \phi_{L1}(0_{+})\big) \nonumber\\
&&~~~~~~~~~~~~~~~~~~~~~~~~~~~~-\frac{r_{k_{1}}}{(k_{1}+\lambda_{\mp})} -\frac{r_{k_{2}}}{(k_{2}+\lambda_{\mp})} \Big), \nonumber\\
c_{R_{\mp}} &=& \pm A\Big(\frac{2\pi}{V_{1}} \big( M_{\pm} \phi_{R2}(0_{+}) - \phi_{R1}(0_{+})\big) \nonumber\\
&&~~~~~~~~~~~~~~~~~~~~~~~~~~~~-\frac{t_{k_{1}}}{(k_{1}+\lambda_{\mp})} -\frac{t_{k_{2}}}{(k_{2}+\lambda_{\mp})} \Big) . \nonumber
\end{eqnarray}

Here, $c_{L_{\mp}}=0$, and $c_{R_{\mp}}=0$ when $U_{1}=0$ and $U_{2}=0$, so that
$\phi_{L1}(x>0), \phi_{L2}(x>0), \phi_{R1}(y>0), \phi_{R2}(y>0)$ have only single-photon behaviours.

%%%%%%%%%%%%%%%%%%%%%%%%%%%%%%%%%%%%%%%%%%%%%%%%%%%%%%%%%%%%%%%%%%%%%
Equations (\ref{twodis1})-(\ref{twodis6}) can be rewritten as
\begin{eqnarray}
\phi_{LL}(0_{+},x_{2}>0) &=&\frac{1}{\sqrt{2}} \frac{1}{2\pi} \Big(  r_{k_{1}} r_{k_{2}} e^{ik_{1}x_{2}} +r_{k_{2}} r_{k_{1}} e^{ik_{2}x_{2}} \nonumber\\
&&+B_{LL_{-}} e^{-i \lambda_{-} x_{2}}+B_{LL_{+}} e^{-i \lambda_{+} x_{2}} \Big), \label{phill0p}\\
\phi_{LR}(0_{+},y_{1}>0) &=&\frac{1}{2\pi} \Big( t_{k_{1}} r_{k_{2}} e^{i k_{1} y_{1}} +t_{k_{2}} r_{k_{1}} e^{i k_{2} y_{1}} \nonumber\\
&&+B_{LR1_{-}}e^{-i\lambda_{-}y_{1}}+B_{LR1_{+}}e^{-i\lambda_{+}y_{1}}\Big),
\label{philr0p}\\
\phi_{RR}(0_{+},y_{2}>0) &=&\frac{1}{\sqrt{2}} \frac{1}{2\pi} \Big( t_{k_{1}} t_{k_{2}} e^{ik_{1} y_{2}} + t_{k_{2}} t_{k_{1}} e^{ik_{2}y_{2}} \nonumber\\
&&+B_{RR_{-}}e^{-i\lambda_{-}y_{2}}+B_{RR_{+}}e^{-i\lambda_{+}y_{2}} \Big),
\label{phirr0p}
\end{eqnarray}
with $B_{LL_{-}} = -i \frac{V_{1}}{\sqrt{2}} M_{-} c_{L_{-}} $, $B_{LL_{+}} = -i \frac{V_{1}}{\sqrt{2}} M_{+}c_{L_{+}} $, $B_{LR1_{-}} = -i\frac{V_{1}}{\sqrt{2}}M_{-}c_{R_{-}}$, $B_{LR1_{+}} = -i\frac{V_{1}}{\sqrt{2}}M_{+}c_{R_{+}}$, $B_{RR_{-}}= -i\frac{V_{2}}{\sqrt{2}} c_{R_{-}} $, and $B_{RR_{+}}=-i\frac{V_{2}}{\sqrt{2}} c_{R_{+}} $.

%\begin{eqnarray}
%B_{LL_{-}} &=& -i \frac{V_{1}}{\sqrt{2}} M_{-} c_{L_{-}} \nonumber\\
%B_{LL_{+}} &=& -i \frac{V_{1}}{\sqrt{2}} M_{+}c_{L_{+}} \nonumber\\
%B_{LR1_{-}} &=&  -i\frac{V_{1}}{\sqrt{2}}M_{-}c_{R_{-}} \nonumber\\
%B_{LR1_{+}} &=&  -i\frac{V_{1}}{\sqrt{2}}M_{+}c_{R_{+}} \nonumber\\
%B_{RR_{-}}&=& -i\frac{V_{2}}{\sqrt{2}} c_{R_{-}} \nonumber\\
%B_{RR_{+}}&=&-i\frac{V_{2}}{\sqrt{2}} c_{R_{+}} \nonumber
%\end{eqnarray}

Finally, substituting Eqs.~(\ref{phil1p}), (\ref{phil2p}), (\ref{phir1p}), and (\ref{phir2p}) and then applying the initial conditions Eqs.~(\ref{phill0p}), (\ref{philr0p}), and (\ref{phirr0p}), we solve Eqs.~(\ref{two1})-(\ref{two3}) in region \ding{194}
\begin{eqnarray}
&&\phi_{LL}(0<x_{1}<x_{2}) = \frac{1}{\sqrt{2}}\frac{1}{2\pi}\Big( r_{k_{1}} r_{k_{2}} e^{i k_{2} x_{1} + i k_{1} x_{2}} + r_{k_{2}} r_{k_{1}} e^{i k_{1}x_{1}+i k_{2}x_{2}} \nonumber\\
&&~~~~~~+ B_{LL_{-}} e^{i (k_{1}+k_{2}+\lambda_{-}) x_{1} - i \lambda_{-} x_{2}} + B_{LL_{+}} e^{i (k_{1}+k_{2}+\lambda_{+}) x_{1} - i \lambda_{+} x_{2}} \Big), \nonumber\\ %\label{phillpp1}\\
&&\phi_{LR}(0<x_{1}<y_{1}) =\frac{1}{2\pi} \Big( t_{k_{1}} r_{k_{2}} e^{i k_{2} x_{1} + i k_{1} y_{1}} + t_{k_{2}} r_{k_{1}} e^{i k_{1}x_{1}+i k_{2}y_{1}} \nonumber\\
&&~~~~~~+ B_{LR1_{-}} e^{i (k_{1}+k_{2}+\lambda_{-}) x_{1} - i \lambda_{-} y_{1}} + B_{LR1_{+}} e^{i (k_{1}+k_{2}+\lambda_{+}) x_{1} - i \lambda_{+} y_{1}} \Big), \nonumber\\ %\label{philrpp1}\\
&&\phi_{RR}(0<y_{1}<y_{2}) = \frac{1}{\sqrt{2}} \frac{1}{2\pi} \Big( t_{k_{1}} t_{k_{2}} e^{i k_{2} y_{1} + i k_{1} y_{2}} + t_{k_{2}} t_{k_{1}} e^{i k_{1}y_{1}+i k_{2}y_{2}} \nonumber\\
&&~~~~~~+ B_{RR_{-}} e^{i (k_{1}+k_{2}+\lambda_{-}) y_{1} - i \lambda_{-} y_{2}} + B_{RR_{+}} e^{i (k_{1}+k_{2}+\lambda_{+}) y_{1} - i \lambda_{+} y_{2}} \Big). \nonumber %\label{phirrpp1}.
\end{eqnarray}

%%%%%%%%%%%%%%%%%%%%%%%%%%%%%%%%%%%%%%%%%%%%%%%%%%%%%%%%%%%%%%%%%%%%%
%Extending these solutions from the half-space to the full space gives rise to the two-photon scattering eigenstate functions given in Eq.(\ref{phiLL})-(\ref{phiRR}) in the main text, where
One can repeat the above calculations for the other half-spaces to obtain
\begin{eqnarray}
\phi_{LL}(0<x_{2}<x_{1}) &=& \phi_{LL}(0<x_{1}<x_{2})\vert_{x_{1} \leftrightarrow x_{2}}, \nonumber\\
\phi_{LR}(0<y_{1}<x_{1}) &=& \frac{1}{2\pi} \Big( r_{k_{1}} t_{k_{2}} e^{i k_{1} x_{1} + i k_{2} y_{1}} +r_{k_{2}} t_{k_{1}} e^{i k_{2}x_{1}+i k_{1}y_{1}} \nonumber\\
&&~~~~~~~~+ B_{LR2_{-}} e^{i (k_{1}+k_{2}+\lambda_{-}) y_{1} - i \lambda_{-} x_{1}} \nonumber\\
&&~~~~~~~~~~~~~~+ B_{LR2_{+}} e^{i (k_{1}+k_{2}+\lambda_{+}) y_{1} - i \lambda_{+} x_{1}} \Big), \nonumber\\
\phi_{RR}(0<y_{2}<y_{1}) &=&\phi_{RR}(0<y_{1}<y_{2})\vert_{y_{1} \leftrightarrow y_{2}},  \nonumber
\end{eqnarray}
where
$B_{LR2_{-}} = -i\frac{V_{2}}{\sqrt{2}}c_{L_{-}}$ and $B_{LR2_{+}} = -i\frac{V_{2}}{\sqrt{2}}c_{L_{+}}$.

From the above results, the full solution of the two-photon eigenstates are given by the amplitudes
\begin{widetext}
\begin{eqnarray}
\phi_{LL}(x_{1},x_{2})&=& 
\frac{1}{\sqrt{2}}\frac{1}{2\pi} \Big(  \sum_{P} \big( \theta(-x_{1})\theta(-x_{2}) + \theta(x_{1})\theta(x_{2}) r_{k_{P_{1}}} r_{k_{P_{2}}} \big) e^{i k_{P_{2}} x_{1} + i k_{P_{1}} x_{2}} \nonumber\\
&&~~~~~~~~~~~~~+ \sum_{Q} e^{i (k_{1}+k_{2})x_{Q_{1}}} \Big( B_{LL_{-}} e^{i \lambda_{-} (x_{Q_{1}}- x_{Q_{2}})} + B_{LL_{+}} e^{i \lambda_{+}( x_{Q_{1}}-x_{Q_{2}})} \Big) \theta(x_{Q_{2}}-x_{Q_{1}})\theta(x_{Q_{1}}) \Big), \label{phiLL}\\
\phi_{LR}(x_{1},y_{1})&=&
\frac{1}{2\pi} \Big( \sum_{P} \theta(x_{1})\theta(y_{1})t_{k_{P_{1}}}r_{k_{P_{2}}} e^{i k_{P_{2}}x_{1}+i k_{P_{1}}y_{1}} \nonumber\\
&&~~~~~~~~~~~~~+e^{i(k_{1}+k_{2})x_{1}} \Big( B_{LR1_{-}} e^{i \lambda_{-} (x_{1} - y_{1})} + B_{LR1_{+}} e^{i \lambda_{+} ( x_{1} - y_{1})}\Big) \theta(y_{1}-x_{1})\theta(x_{1})\nonumber\\
&&~~~~~~~~~~~~~~~~~~~~~~~~~~~~~~~~~~~~~~~~
+ e^{i(k_{1}+k_{2})y_{1}} \Big(B_{LR2_{-}} e^{i \lambda_{-}( y_{1} -x_{1})}+ B_{LR2_{+}} e^{i \lambda_{+} (y_{1} -x_{1})} \Big) \theta(x_{1}-y_{1})\theta(y_{1}) \Big), \label{phiLR} \\
\phi_{RR}(y_{1},y_{2})&=& 
\frac{1}{\sqrt{2}} \frac{1}{2\pi} \Big( \sum_{P} \theta(y_{1})\theta(y_{2}) t_{k_{P_{1}}}t_{k_{P_{2}}} e^{i k_{P_{2}} y_{1} + i k_{P_{1}} y_{2}} \nonumber\\
&&~~~~~~~~~~~~~+ \sum_{Q} e^{i(k_{1}+k_{2})y_{Q_{1}}}\Big(  B_{RR_{-}} e^{i \lambda_{-} (y_{Q_{1}} - y_{Q_{2}})} + B_{RR_{+}} e^{i \lambda_{+} (y_{Q_{1}} - y_{Q_{2}})} \Big)\theta(y_{Q_{2}}-y_{Q_{1}})\theta(y_{Q_{1}}) \Big), \label{phiRR}
\end{eqnarray}
\end{widetext}
where $E^{(2)}=k_1+k_2$.
$P=(P_{1},P_{2})$ and $Q=(Q_{1},Q_{2})$ are permutations of $(1,2)$ needed to account for the bosonic symmetry of the wave function. 

  Here, all the $B$'s become zero if the cavities are linear, i.e., $U_{1}=U_{2}=0$, so that each photon undergoes the individual scattering process and the energy of each photon is preserved. If the system, on the other hand, is nonlinear, the bound-state contributions become important, modifying the photon statistics of the output light as shown in the main text. 
In the limit of $U\rightarrow \infty$, $B$'s become exactly the same as those of the coupled two-level atoms \cite{Roy13}.
Finally, we can find the two-photon scattering matrix from, as in \cite{Zheng10},
\begin{eqnarray}
{\bf S}^{(2)}=\int dk_{1} dk_{2} \frac{1}{2{\rm !}} \ket{\phi_{\rm out}^{(2)}}_{k_{1}, k_{2}}\bra{\phi_{\rm in}^{(2)}}, 
\label{scatteringmatrix2}
\end{eqnarray}
where the input and output states are written as 
\begin{eqnarray}
\ket{\phi_{\rm in}^{(2)}}_{k_{1},k_{2}}&=&\int dx_{1}dx_{2} \phi_{LL}(x_{1}{\rm<}0,x_{2}{\rm<}0) \frac{1}{\sqrt{2}} \opd{c}{L}(x_{1}) \opd{c}{L}(x_{2})\ket{0}\nonumber\\
\ket{\phi_{\rm out}^{(2)}}_{k_{1},k_{2}}&=&\ket{\phi_{\rm out}^{(2)}}_{LL} + \ket{\phi_{\rm out}^{(2)}}_{LR} + \ket{\phi_{\rm out}^{(2)}}_{RR}, \nonumber
\end{eqnarray}
where 
\begin{eqnarray}
\ket{\phi_{\rm out}^{(2)}}_{LL}&=&\int dx_{1}dx_{2} \phi_{LL}(x_{1}{\rm>}0,x_{2}{\rm>}0) \frac{1}{\sqrt{2}} \opd{c}{L}(x_{1}) \opd{c}{L}(x_{2})\ket{0}, \nonumber\\
\ket{\phi_{\rm out}^{(2)}}_{LR}&=&\int dx_{1}dy_{1} \phi_{LR}(x_{1}{\rm>}0,y_{1}{\rm>}0)\opd{c}{L}(x_{1}) \opd{c}{R}(y_{1})\ket{0},\nonumber\\ 
\ket{\phi_{\rm out}^{(2)}}_{RR}&=&\int dy_{1}dy_{2} \phi_{RR}(y_{1}{\rm>}0,y_{2}{\rm>}0) \frac{1}{\sqrt{2}} \opd{c}{R}(y_{1}) \opd{c}{R}(y_{2})\ket{0}.\nonumber
\end{eqnarray}

The scattering matrix elements between the input ($k_{1},k_{2}$) and output ($p_{1},p_{2}$) momentums are given as
\begin{widetext}
\begin{eqnarray}
&& _{LL}\bra{p_{1},p_{2}} {\bf S}^{(2)} \ket{k_{1},k_{2}} = r_{k_{1}} r_{k_{2}} \delta(k_{1}-p_{1})\delta(k_{2}-p_{2}) + r_{k_{2}} r_{k_{1}} \delta(k_{1}-p_{2})\delta(k_{2}-p_{1})
+S_{LL}  \delta(k_{1}+k_{2}-p_{1}-p_{2}), \label{App_scatering_matrix_element_LL}  \\
&& _{LR}\bra{p_{1},p_{2}} {\bf S}^{(2)} \ket{k_{1},k_{2}}= r_{k_{1}} t_{k_{2}} \delta(k_{1}-p_{1})\delta(k_{2}-p_{2}) + r_{k_{2}} t_{k_{1}} \delta(k_{1}-p_{2})\delta(k_{2}-p_{1})
+S_{LR}  \delta(k_{1}+k_{2}-p_{1}-p_{2}),\label{App_scatering_matrix_element_LR}\\
&& _{RR}\bra{p_{1},p_{2}} {\bf S}^{(2)} \ket{k_{1},k_{2}} = t_{k_{1}} t_{k_{2}} \delta(k_{1}-p_{1})\delta(k_{2}-p_{2}) +  t_{k_{2}} t_{k_{1}}  \delta(k_{1}-p_{2})\delta(k_{2}-p_{1})
+S_{RR}  \delta(k_{1}+k_{2}-p_{1}-p_{2}),\label{App_scatering_matrix_element_RR}
\end{eqnarray}
where
\begin{eqnarray}
S_{LL} &=&  \frac{1}{2\pi} \Big( 
B_{LL_{-}}^{k_{1},k_{2}} ( \frac{-i}{ \lambda_{-}^{k_{1},k_{2}}+p_{2}} +\frac{-i}{\lambda_{-}^{k_{1},k_{2}}+p_{1}} ) 
+ B_{LL_{+}}^{k_{1},k_{2}} ( \frac{-i}{ \lambda_{+}^{k_{1},k_{2}}+p_{2}} + \frac{-i}{\lambda_{+}^{k_{1},k_{2}}+p_{1}})
 \Big), \nonumber\\
 S_{LR} &=& \frac{1}{2\pi} \Big( 
B_{LR1_{-}}^{k_{1},k_{2}}  \frac{-i}{ \lambda_{-}^{k_{1},k_{2}}+p_{2}} 
+B_{LR1_{+}}^{k_{1},k_{2}}  \frac{-i}{ \lambda_{+}^{k_{1},k_{2}}+p_{2}} 
+B_{LR2_{-}}^{k_{1},k_{2}}  \frac{-i}{ \lambda_{-}^{k_{1},k_{2}}+p_{1}}
+B_{LR2_{+}}^{k_{1},k_{2}}\frac{-i}{ \lambda_{+}^{k_{1},k_{2}}+p_{1}}
 \Big), \nonumber\\
 S_{RR} &=&  \frac{1}{2\pi} \Big( 
B_{RR_{-}}^{k_{1},k_{2}} ( \frac{-i}{ \lambda_{-}^{k_{1},k_{2}}+p_{2}} +\frac{-i}{\lambda_{-}^{k_{1},k_{2}}+p_{1}} ) 
+ B_{RR_{+}}^{k_{1},k_{2}} ( \frac{-i}{ \lambda_{+}^{k_{1},k_{2}}+p_{2}} + \frac{-i}{\lambda_{+}^{k_{1},k_{2}}+p_{1}})
 \Big). \nonumber
\end{eqnarray}
\end{widetext}

%%%%%%%%%%%%%%%%%%%%%%%%%%%%%%%%%%%%%%%%%%%%%%%%%%%%%%%%%%%%%%%%%%%%%
\section{Intensity-intensity correlation}

In this section, we discuss the equal-time second-order intensity correlations of the initial two-photon wavepacket and study the effects of pulse-shape on the correlations of the transmitted light. 

%%%%%%%%%%%%%%%%%%%%%%%%%%%%%%%%%%%%%%%%%%%%%%%%%%%%%%%%%%%%%%%%%%%%%

\subsection{Correlations between the two initial photons}\label{App_initial_g2}
\begin{figure}[b]
\centering
\includegraphics[width=8cm]{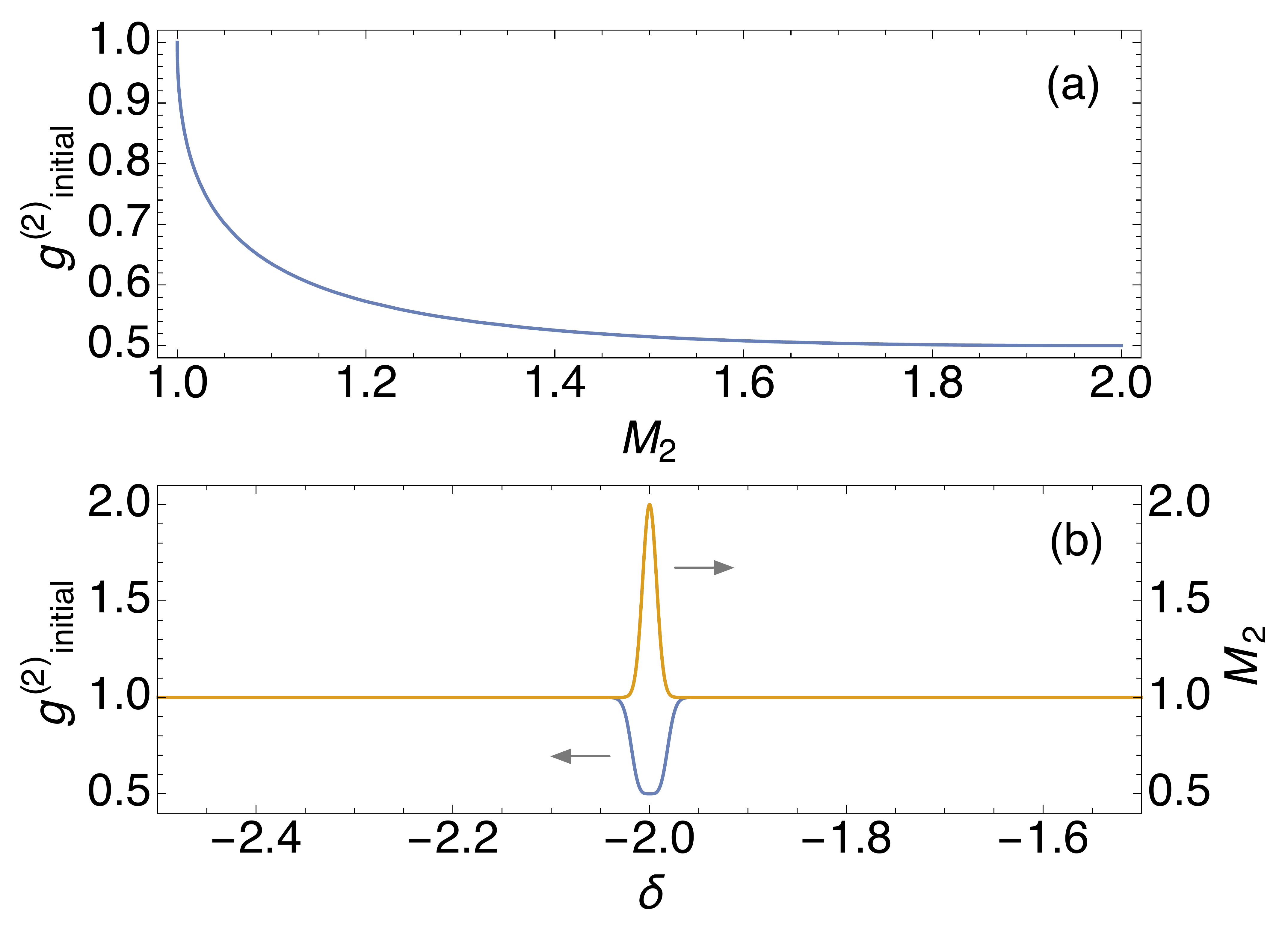}
\caption{ 
(a) $g^{(2)}_{\rm initial}(0)$ as a function of $M_{2}$, constructed from different values of momentums $k_{1}$ and $k_{2}$ for initial two photons. 
(b) $g^{(2)}_{\rm initial}$ as a function of $\delta$ when $\Delta k = \delta - 2\epsilon_{-}^{(1)}$.
}
\label{SMinitialg2function} 
\end{figure}

Here, we analyse the initial correlations for two photons given in the main text: $\ket{2_{\{ \xi \}}} = \frac{1}{\sqrt{M_{2}}} \opd{c}{\xi_{1}}\opd{c}{\xi_{2}}\ket{0}$. The correlation function $g^{(2)}_{\rm initial}(x_{1},x_{2})$ can be written as
\begin{equation}
g^{(2)}_{\rm initial} (x_{1},x_{2})= \frac{
\vert g_{1}(x_{1})g_{2}(x_{2})+g_{1}(x_{2})g_{2}(x_{1})\vert^{2}
}
{
\frac{1}{M_{2}}
g_{3}(x_{1}) g_{3}(x_{2})
},\nonumber
\end{equation}
where
\begin{eqnarray}
g_{1}(x) &=& \frac{1}{\sqrt{2\pi}} \int dk \xi_{1}(x) e^{i k x}, \nonumber\\
g_{2}(x) &=& \frac{1}{\sqrt{2\pi}} \int dk \xi_{2}(x) e^{i k x}, \nonumber\\
g_{3}(x) &=&  \vert g_{1}(x) \vert^{2} +\vert g_{2}(x) \vert^{2}\nonumber\\
&&~~~~~~~ +\sqrt{M_{2}-1} \big(g_{1}(x) g_{2}^{*}(x) +  g_{2}(x) g_{1}^{*}(x) \big). \nonumber
\end{eqnarray}
In Fig.~\ref{SMinitialg2function}(a),
we depict a monotonic relation between the auto-correlation, $g^{(2)}_{\rm initial}(0)$, and the overlap of initial wave packets, $M_{2}$, constructed from different values of momenta $k_{1}$ and $k_{2}$. The auto-correlation $g^{(2)}_{\rm initial}(0)$ has a maximum at $M_{2}=1$ (corresponding to when $\abs{k_{1}-k_{2}}\gg\sigma$) and a minimum at $M_{2}=2$ (corresponding to when $k_{1}=k_{2}$).
Figure~\ref{SMinitialg2function}(b) shows that $g^{(2)}_{\rm initial}$ has a minimum of $0.5$ at $\delta=2\epsilon_{-}^{(1)}$ when $\Delta k = \delta - 2\epsilon_{-}^{(1)}$ (corresponding to the case of $\Delta k=0$), i.e., $g^{(2)}_{\rm initial}$ depends on $\delta$ when $\Delta k = \delta - 2\epsilon_{-}^{(1)}$, while $g^{(2)}_{\rm initial}=0.5$ when $\Delta k =0$ regardless of $\delta$.

%%%%%%%%%%%%%%%%%%%%%%%%%%%%%%%%%%%%%%%%%%%%%%%%%%%%%%%%%%%%%%%%%%%%%
\subsection{Effects of pulse shape in narrow-band regime}\label{App_pulse_shape}
\begin{figure}[b]
\centering
\includegraphics[width=8cm]{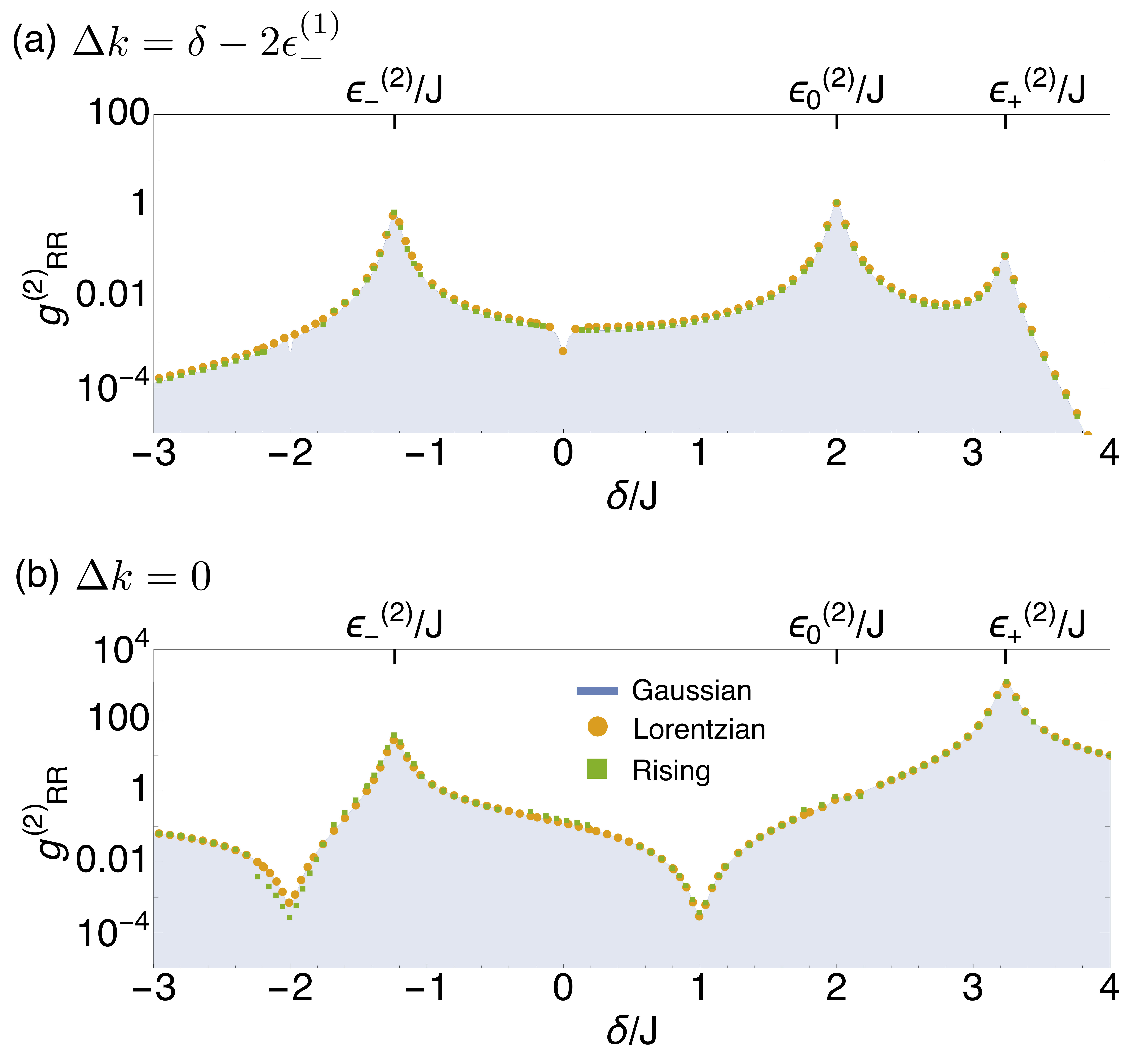}
\caption{$g_{RR}^{(2)}$ as a function of detuning $\delta/J$ for $U/J=1$ when $\Delta k = \delta - \epsilon_{-}^{(1)}$ in (a) and $\Delta k=0$ in (b). Parameters are the same as in the main text. }
\label{SMg2pulseshape}
\end{figure}

In this section, we show that the effects of pulse shape in photon scattering is negligible given a narrow enough bandwidth. For this purpose, we examine equal-time auto-correlations in the transmitted light, $g_{RR}^{(2)}$, for three different temporal envelopes, Gaussian, Lorentzian, and Rising distributions, respectively given as
\begin{eqnarray}
\Xi_{\rm G}(t) &=& \sqrt{\sigma} {\rm exp}[- \sigma^{2}t^{2} -i k_{0}t] (2/\pi)^{1/4}, \nonumber \\
\Xi_{\rm L}(t) &=& \sqrt{\sigma} {\rm exp}[-\sigma \vert t \vert -i k_{0}t ], \nonumber \\
\Xi_{\rm R}(t) &=& \sqrt{\sigma} {\rm exp}[\sigma t /2 -i k_{0}t]\theta(-t),\nonumber
\end{eqnarray}
where $\sigma$ is the inverse temporal pulse width and $k_{0}$ is the central momentum. 
In the momentum space, they read
\begin{eqnarray}
\xi_{\rm G}(k) &=& {\rm exp}[-(k-k_{0})^{2}/4\sigma^{2}](2\pi \sigma^{2})^{-1/4}, \nonumber \\
\xi_{\rm L}(k) &=& \sqrt{2/\pi} \sigma^{3/2}((k-k_{0})^{2}+\sigma^{2})^{-1}, \nonumber \\
\xi_{\rm R}(k) &=&\sqrt{2/\pi} \sqrt{\sigma}(2i (k-k_{0})+\sigma)^{-1}, \nonumber 
\end{eqnarray}
where $\sigma$ can be seen as the bandwidth of each profile.

Figure~\ref{SMg2pulseshape} plots $g_{RR}^{(2)}$ as a function of the probe detuning for three different pulse profiles. The continuous lines are the results for the Gaussian profile, whereas the results for the Lorentzian and Rising profiles are marked by the red and blue dots, respectively. These results clearly demonstrate the insensitivity of the intensity correlations to the pulse profile, as expected in the narrow-band regime. We have also checked that the probabilities are similarly insensitive to the pulse profile.

%%%%%%%%%%%%%%%%%%%%%%%%%%%%%%%%%%%%%%%%%%%%%%%%%%%%%%%%%%%%%%%%%%%%%

%From Eqs.~(\ref{App_g2RR})-(\ref{App_Heff}), 
%%%%%%%%%%%%%%%%%%%%%%%%%%%%%%%%%%%%%%%%%%%%%%%%%%%%%%%%%%%%%%%%%%%%%
%%%%%%%%%%%%%%%%%%%%%%%%%%%%%%%%%%%%%%%%%%%%%%%%%%%%%%%%%%%%%%%%%%%%%


\begin{thebibliography}{99}
\bibitem{footnote} We note here that many-body spectroscopy in a different context has also been used to probe quantum magnetism in  \cite{Kurcz14}.
\end{thebibliography}
\end{document}